# Intermediate stages in the origin of metabolism at a phosphorylating hydrothermal vent

Short Title: Metabolic network origin at a hydrothermal vent

Natalia Mrnjavac[1,†,*], Nadja K. Hoffmann[1,†], Manon L. Schlikker[1,†], Maximilian Burmeister[1], Loraine Schwander[1], Carolina García García[1], Max Brabender[1], Mike Steel[2], Daniel H. Huson[3], Sabine Metzger[1], Quentin Dherbassy[4], Bernhard Schink[5], Mirko Basen[6], Joseph Moran[4,7], Harun Tüysüz[8,9], Martina Preiner[10], William F. Martin[1]

[1] Institute of Molecular Evolution, Faculty of Mathematics and Natural Sciences, Heinrich Heine University Düsseldorf, Düsseldorf, Germany

[2] Biomathematics Research Centre, University of Canterbury, Christchurch, New Zealand

[3] Institute for Bioinformatics and Medical Informatics, University of Tübingen, Tübingen, Germany

[4] University of Strasbourg, CNRS, ISIS, Strasbourg, France

[5] Department of Biology, University of Konstanz, Konstanz, Germany

[6] Department of Microbiology, Institute of Biological Sciences, University of Rostock, Rostock, Germany

[7] Department of Chemistry and Biomolecular Sciences, University of Ottawa, Canada

[8] Catalysis and Energy Materials Group, IMDEA Materials Institute, Madrid, Spain

[9] Department of Heterogeneous Catalysis, Max-Planck-Institut für Kohlenforschung, Mülheim an der Ruhr, Germany

[10] Microcosm Earth Center, Max Planck Institute for Terrestrial Microbiology and Philipps University Marburg, Marburg, Germany

† These authors contributed equally to this work.

* Correspondence: N.Mrnjavac@hhu.de

*Abstract*

The origin of life required the emergence of metabolism, an autocatalytic network of enzymatic reactions that synthesize amino acids, nucleotides and cofactors. At the origin of metabolism there were no enzymes—how did it start? Empirical studies addressing early metabolic evolution are lacking. Harnessing protein structures for metabolic enzymes, we identify intermediate states in primordial metabolic assembly. We show that enzymatic metabolism in the universal common ancestor was incomplete, undergoing final assembly independently in the lineages leading to Bacteria and Archaea. Native transition metals—$Fe^0$, $Co^0$, $Ni^0$, $Pd^0$—served as the catalytic forerunners of both enzymes and cofactors at metabolic origin while phosphite supplied energy, as it phosphorylates AMP to ADP and serine to phosphoserine using native metal catalysts in water. Phosphite and native metals occur in serpentinizing hydrothermal systems, identifying an energy-supplying, catalytic site of metabolic origin. Cofactors liberated nascent metabolism from native metal catalysts, engendering its autocatalytic state.

Teaser: Enzymatic metabolism emerged from metal-catalyzed, aqueous reactions of $H_2$, $CO_2$, $NH_3$ and phosphite at a hydrothermal vent.

## Introduction

Metabolism is a non-negotiable property of life. It transforms environmental compounds into the building blocks of life and the chemical substance of cells. Though modern metabolism is catalyzed by enzymes, at origins there were no enzymes, leaving only the environment as the source of compounds and catalysts that spawned the first metabolic networks. Yet the environment in which life and metabolism arose is hotly debated, suggestions including uv-rich terrestrial hot springs[1], terrestrial cyanide deposits[2], volcanos[3,4], ice[5], and submarine hydrothermal vents[6–8]. Because the environment of origins conditions its chemistry, almost everything about the origin of metabolism is debated, including the roles of energy[9], genetics[10], autocatalysis[11], phosphate[12,13], cofactors[14], cyanide[15], $CO_2$ [3], and water[16,17]. Yet on one aspect all will agree: the ~400-reaction network that converts $H_2$, $CO_2$, $NH_3$, $H_2S$ and phosphate into amino acids, bases and cofactors[18] cannot have arisen in an instant. Its emergence from spontaneous environmental reactions had to traverse intermediate states of assembly, which have previously been elusive.

A key question about metabolic origin concerns its state in the last universal common ancestor, LUCA[19,20]. Was LUCA able to synthesize all of its metabolic intermediates itself or were some still supplied by the environment, and were all of LUCA's reactions catalyzed by enzymes or did the environment supply catalysts that served as the forerunners of LUCA's enzymes? And if the environment supplied compounds and catalysts for LUCA, how far did that dependency extend into the lineages leading to the last common ancestors of bacteria and archaea (LBCA and LACA) respectively? Though the timing of bacterial and archaeal divergence has been extensively studied[21], the processes of enzyme origins and metabolic assembly across the deepest divide of the tree of life have not.

The 400 reactions of metabolism and corresponding enzymes can provide insights into the earliest phases of biological divergence, but they might also harbor evidence for the kind of environment within which metabolism arose. Do reactions and enzymes that trace to LUCA indicate a different kind of chemistry or a different environment from those that trace to LACA and LBCA and can the different theories for the site of life's origin be discriminated on the basis of metabolism's chemistry? Neither question has been explored on the basis of metabolism's conserved 400 reaction set.

Perhaps the steepest barrier to understanding metabolic origin concerns the role of energy and phosphate. Regardless of how or where metabolism arose, its reactions must have been thermodynamically favorable, otherwise they would not have gone forward and life's metabolic network never would have arisen. Central to energetics in metabolism is phosphate[22,23]. Phosphate is poorly soluble and does not readily form reactive anhydrides or organophosphate compounds in water[23], prompting proposals that metabolism either arose in environments that underwent recurrent desiccation, which promotes phosphorylation reactions[23,24] but denatures enzymes, or arose without phosphorus altogether[25]. The letter suggestion seems unlikely, as the conserved reactions of amino acid, nucleotide and cofactor synthesis involve 360 intermediates, 114 of which contain one or more phosphate residues—phosphate has been part of metabolism since there were genes. Compounds capable of phosphorylation had to exist in the environment where metabolism arose, but environmental sources of aqueous phosphorylation have not been identified to date.

To chart the course of early metabolic evolution, we harnessed information in protein structures for metabolic enzymes to identify the enzymes present in the metabolism of LUCA and the growth of that reaction network in the independent lineages leading to Bacteria and Archaea. This identified intermediate states in primordial metabolic assembly in which enzymatic metabolism was incomplete, with missing enzymes and intermediates being supplied by native transition metals—$Fe^0$, $Co^0$, $Ni^0$ and their alloys—which are known to catalyze essential metabolic pathways[26–33] and which are naturally deposited in serpentinizing hydrothermal systems[34–37]. Guided by microbes that utilize phosphite as their sole source of phosphorus[38,39], we found that another native metal of serpentinizing hydrothermal vents[36,37] , $Pd^0$, provided a hitherto missing source of phosphorylation at origins, as it catalyzed phosphite-dependent phosphorylation of AMP and serine in water. Phosphite also occurs in serpentinizing systems[12], its activation to phosphate in water is highly exergonic[23], identifying a natural supply of continuous chemical energy and implicating serpentinizing hydrothermal vents as the site of metabolic assembly.

## Results and discussion

### *Core metabolism in LUCA was incomplete*

To identify intermediate states in the primordial assembly of metabolism, we harnessed information available in enzyme sequences and structures, which can detect more ancient homologies among proteins than sequence comparisons alone[40]. Starting from a balanced set of genomes encompassing 552 bacterial and 401 archaeal isolates with representation in all major groups, and excluding metagenomic assembled genomes, we generated structure-refined clusters of enzymes that map to 361 reactions required to synthesize amino acids, bases, and cofactors from gasses ($H_2$, $CO_2$, $NH_3$, $H_2S$) and mineral salts (see **Methods**). The standard Markov Cluster Algorithm (MCL) sequence clustering procedure initially generated 1473 clusters, which were merged by structural similarity, yielding 426 structurally refined clusters (**Fig. S1**).

The diversity of theories for metabolic origin[3,9–11,15,26–30,41–49]generates few predictions for the expected evolutionary trajectory of enzymatic metabolism. The ribosome[50,51] can, however, serve as a predictive model for the emergence of complex traits. The ribosome of LUCA[50,51] was simpler than its modern forms (**Fig. 1A)**, containing only 33 ribosomal proteins. It underwent independent lineage-specific assembly via addition of 29 and 21 novel proteins in lineages leading to LACA and LBCA, respectively[50,51] (**Fig. 1A, Table S1)**. We asked whether the assembly of metabolism, which supplies the RNA and amino acid monomers required for ribosome synthesis and function, proceeded via a ribosome-like trajectory from a simple intermediate state in LUCA, followed by lineage-specific addition of novel reactions.

Plotting the phylogenetic distributions of core biosynthetic enzymes across bacteria and archaea reveals that enzymatic metabolism in LUCA was incomplete (**Fig. 1B, Table S2)**. It expanded via origins of novel enzymes in the lineages leading to LACA and LBCA, closely mirroring lineage-specific assembly of the ribosome[50,51]. In the present sample, 166 enzymes of core metabolism trace to LUCA, yet 89 enzymes required for the synthesis of amino acids, cofactors and bases arose on the lineage to LBCA, while 38 arose on the lineage to LACA. An additional 37 enzymes were too sparsely distributed for unequivocal lineage attribution (gray shading in **Fig. 1B**). These post-LUCA bacterial and archaeal enzyme innovations reveal that the insufficiency of enzymatic metabolism in LUCA was severe and provide novel insights into an extremely early phase of biochemical evolution. The metabolic reactions of LUCA (**Fig. 1B)** capture a time in which the ribosome, the genetic code and translation were functional, because enzymes existed and were arising *de novo*. Yet enzymes were evolving without the support of a complete enzymatic supply of precursors for protein synthesis: Amino acid synthesis, cofactor synthesis and intermediate carbon metabolism were incomplete in LUCA (**Fig. 2**). How could an incomplete enzymatic metabolism support protein evolution at the ribosome?

The principle of autocatalysis offers one avenue of explanation in that catalysts can generate products that generate new catalysts such that metabolic networks can grow[11,49]. Yet the problem at the origin of metabolism is not whether autocatalytic networks can arise in nature—they can[52]—the problem is practical: Where do sufficient concentrations of the very first amino acids and nucleobases come from that permit the synthesis of ribosomes for the synthesis of metabolic enzymes that set the reactions of life moving forward? Two broad schools of thought exist on the primordial supply of monomers at metabolic origins.

One proposal is that a primordial, cyanide-dependent chemistry disjunct from modern, $CO_2$-dependent metabolism generated the first building blocks of life[15,53]. A vast number of prebiotic reactions have been reported that synthesize essential biomolecules, yet using reactants that do not occur in metabolism[15,53,54]. These traditioned laboratory syntheses clearly demonstrate that biological compounds can arise by varied and elegant abiotic routes, in addition to biosynthetic routes, but they do not speak to the origin or evolution of enzymatic reactions of metabolism as they occur in cells (**Fig. 1B, Table S2**), which at the level of ecosystems always start from $CO_2$[55,56]. Cyanide-dependent synthesis thus requires two origins of metabolism, one from cyanide and the modern version from $CO_2$. An alternative to this two-origin proposal is that naturally existing catalysts, if provided with a continuous environmental source of simple biochemical reactants ($H_2$, $CO_2$, $NH_3$), promoted $CO_2$-based reactions virtually identical to those of modern metabolism[26,31,32,44,45,57], and that catalytic activities present in the environment where metabolism arose were later replaced by enzymes and cofactors[42–45,58,59] , allowing the process of metabolic evolution to go forward, both in LUCA and its descendant lineages. This would require only one origin of metabolism, with a continuity of chemical reactions from environment to cells, accompanied however by an increasingly complex set of catalysts (enzymes and cofactors), synthesized by metabolism with the effect of reinforcing and accelerating network formation[52]. To distinguish between these possibilities, we investigated the reactions of metabolism.

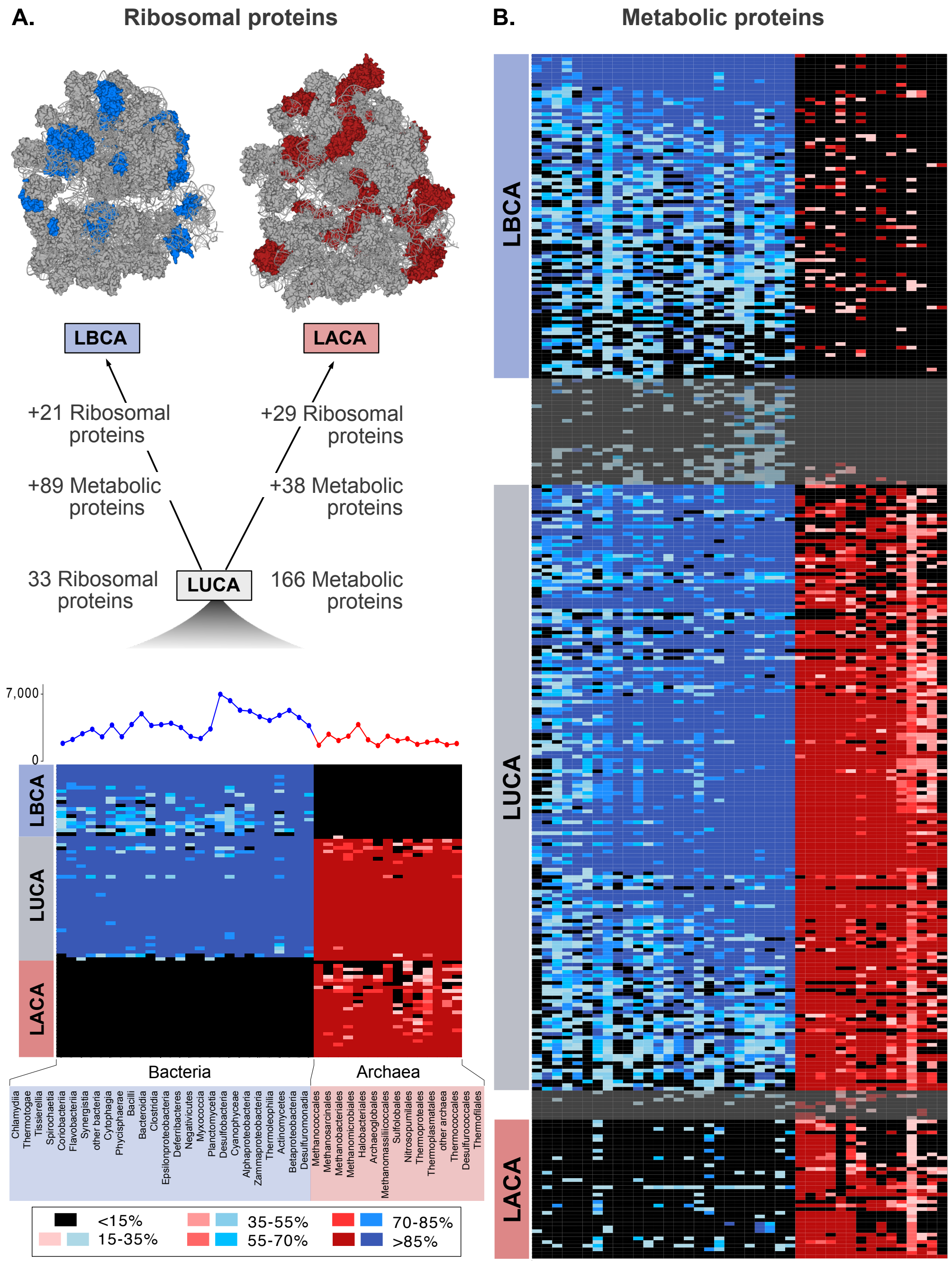

**Fig. 1. Distribution of ribosomal and core metabolism protein families across bacterial classes and archaeal orders.** Protein families in bacterial genomes are plotted as blue ticks, archaeal in red, color intensity indicates the proportion of genomes per taxon possessing the gene (scale at lower left). **A.** Universal ribosomal proteins (tracing to LUCA) are gray in the ribosome structures, bacterial-specific ribosomal proteins shown in blue, and archaeal-specific in red. The schematic tree shows the origin of ribosomal proteins[50,51] and metabolic proteins (this paper). The line graph shows the average genome size (proteins) per taxon. The distribution of LUCA, Bacteria- and Archaea-specific ribosomal proteins displays a characteristic pattern. Taxa are indicated in the lower panel. **B.** The distribution of core metabolic protein families mirrors

ribosomal protein origin in **A**. Protein families clearly attributable to LUCA, LBCA (top) and LACA (bottom) (see **Methods**), are marked. Gray zones designate protein families that are sparsely distributed and hence difficult to assign to LUCA or lineages (see **Methods)**. The probability of observing, by chance, the dichotomy of columns along the bacterial vs. archaeal split in **B** is 0.002 (two-sided permutation test).

***Metabolism harbors evidence for the site of its origin***

In both **Fig. 1B** and **Fig. 2**, the metabolic reactions enzymatically catalyzed in LUCA uncover an intermediate state during metabolic origin that underwent expansion in its descendant lineages. From this intermediate state a salient observation emerges: While missing enzymes for reactions in LUCA's metabolism were still evolving, the initial network remained intact in the lineages leading to both LACA and LBCA, as did the initial inputs to the network: $H_2$, $CO_2$, $NH_3$, $H_2S$, $H_2O$ and inorganic phosphate. This unconditionally requires continuity[45] in the process of metabolic evolution and its environment, bearing out, a century after its formulation, Kluyver's[60] central postulate that there is unity in biochemistry, also during its origin. The growth of metabolism reveals continuity during the course of metabolic assembly from LUCA to LACA and LBCA. In much the same way as the genetic code and the ribosome provide strong evidence for a single origin of life[50,51], the metabolic reactions of LUCA reveal an intermediate state in metabolic evolution (**Fig. 1B**) that provides strong evidence for a single, continuously reactive chemical environment as the site of metabolic origin.

But what kind of chemical environment? Again, the reactions of metabolism speak to this question because they identify a starting point: $CO_2$ reduction to pyruvate via the acetyl-CoA pathway. It is a linear pathway of $H_2$-dependent $CO_2$ reduction[55,58] and is the only one that operates in both bacteria and archaea[55,56]. It uses Fe, Ni and Co at the active site of its enzymes for $H_2$ activation and $CO_2$ conversion to acetyl-CoA[61–63]. It energetically couples $H_2$-dependent $CO_2$ fixation to ATP synthesis in acetogens (bacteria) and methanogens (archaea)[9]. In phylogenetic reconstructions, it traces to LUCA[19,20], to LACA[64] and LBCA[65], albeit requiring different cofactors[66] and unrelated enzymes[67] in the archaeal and bacterial methyl synthesis branch. Moreover, under conditions of modern serpentinizing hydrothermal systems (alkaline aqueous solution, $H_2$ partial pressures on the order of 5 atm), catalysts that are naturally deposited in serpentinizing systems—$Ni^0$, $Fe^0$, $Ni_3Fe$[34] and $Fe_3O_4$[35,68]—convert $H_2$ and $CO_2$ in the laboratory overnight at 20–100°C in water to formate, acetate and pyruvate (plus methane), the intermediates and end products of the acetyl-CoA pathway[26,28,31]. Serpentinizing hydrothermal vents still generate $H_2$[69], native metals[34], abiotic formate and methane[69,70] today. Among known biochemical pathways, such congruence between geochemically catalyzed and enzymatic multistep reactions has no precedent. This anchors the origin of metabolism to the acetyl-CoA pathway via transition metal-catalyzed reactions of $H_2$ and $CO_2$ on mineral surfaces of serpentinizing hydrothermal systems[57], environments that have continuously existed on Earth since the existence of liquid water[35].

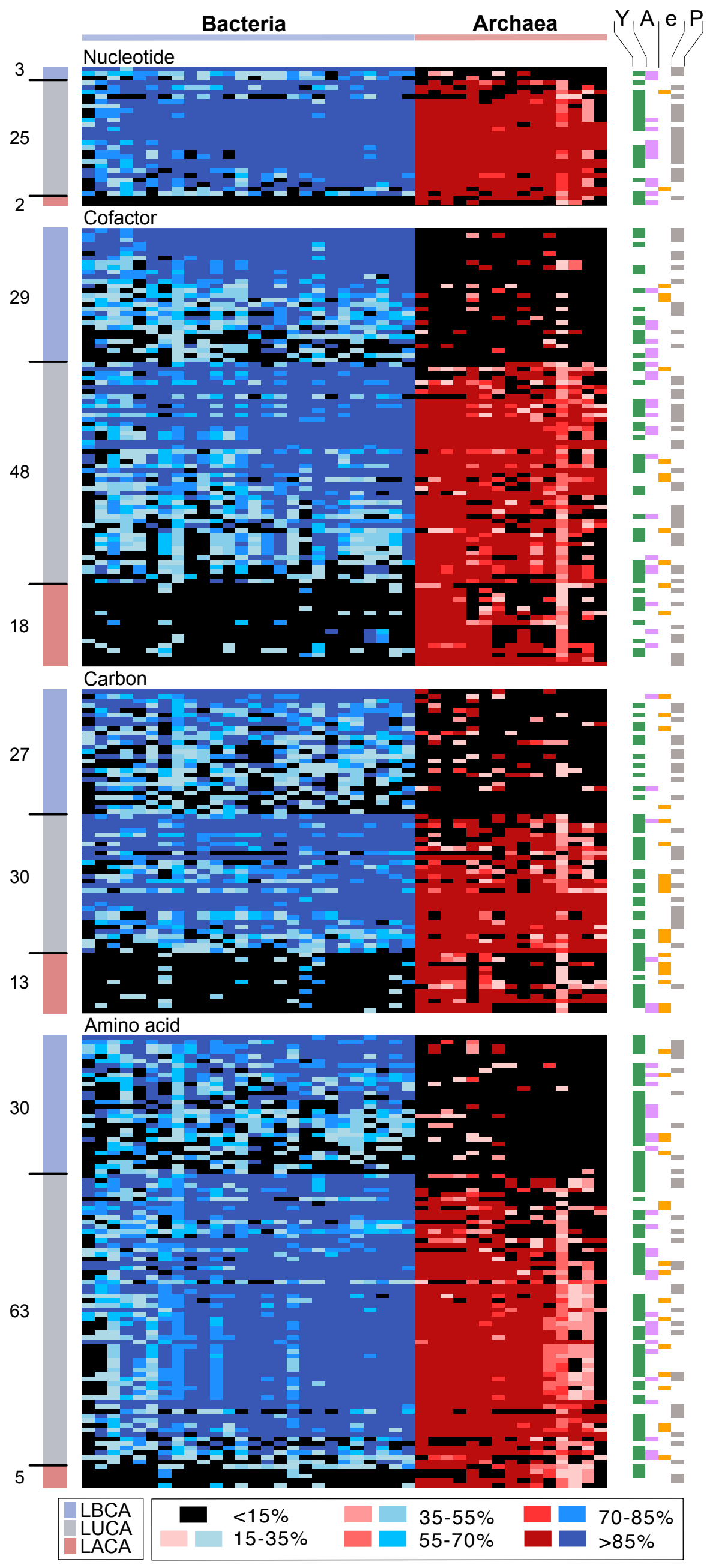

**Fig. 2. Distribution of core metabolism protein families by sectors of metabolism.** The rows are identical to those in **Fig. 1B**, but grouped by function. Gray areas from **Fig. 1B** are excluded. The number of protein families per category is shown on the left. Columns on the right denote reactions that involve carbonyl groups (Y, green), C-N bond formation (A, pink), electron transfer (e, yellow) and formation or cleavage of phosphate bonds (P, gray), as listed in **Table S3**.

Among the four main sectors of metabolism—nucleotide synthesis, cofactors synthesis, amino acid synthesis and central carbon metabolism—nucleotide synthesis stands out as enzymatically complete in LUCA (**Fig. 2**). This reflects its irreplaceable nature. The main function of nucleotides in cells is not small molecule synthesis, but transmission and expression of genetic information, functions which cannot be replaced by minerals and water. Several proposals for surface-catalyzed reactions that mirror biochemical nucleotide synthesis suggest a simple chemistr[3,45], yet involving ionic bonds of substrates to surfaces rather than the covalent carbon-metal bonds formed by native transition metals[26,31]. By contrast, amino acid synthesis was incomplete in LUCA, requiring catalytic input from the environment and enzyme inventions in the lineages leading to LACA and LBCA (**Fig. 2**). In metabolism, amino acids are synthesized from 2-oxoacids via reductive aminations or transaminations[71]. The same reactions are efficiently catalyzed by $Ni^0$[29], as are rTCA cycle reactions that generate biochemically relevant 2-oxoacid precursors from pyruvate and glyoxylate[27,29,72]. Such non-enzymatic reactions permitted continuous synthesis of simple amino acids, in particular ancient ones[73], during the assembly of metabolism prior to the origin of translation.

***Metabolic assembly uncovers independent enzyme origins***

The finding that metabolism underwent final assembly post-LUCA (**Fig. 2**) generates the prediction that independent origins of unrelated enzymes for the same reaction should be identifiable. We detected five such cases in which bacteria and archaea invented structurally unrelated enzymes to catalyze the same reaction, without the gene subsequently having undergone rampant transdomain transfer[19,20] subsequent to its origin. These examples are shown in **Fig. 3**. Alanine dehydrogenase, AlaDH (EC 1.4.1.1) (**Fig. 3E**) catalyzes the reversible, NADH-dependent reductive amination of pyruvate to alanine[74]. Independent origins of AlaDH in the LBCA and LACA lineages indicates that the environment provided this chemically facile function[29,30] in LUCA. Two enzymes of the shikimate pathway document post-LUCA innovations (**Fig. 3**): shikimate kinase and 3-dehydroquinate dehydratase. Archaea synthesize shikimate from different precursors than bacteria, though the shikimate kinase (**Fig. 3C**) and 3-dehydroquinate dehydratase (**Fig. 3D**) reactions are conserved[75], yet catalyzed by unrelated enzymes, suggesting a late enzymatic origin of aromatic amino acid synthesis[73]. The shikimate pathway also leads to pterins, including tetrahydrofolate ($H_4F$) and tetrahydromethanopterin ($H_4MPT$), the methyl carriers in the acetyl-CoA pathway of bacteria and archaea. Also required for $H_4F$ and $H_4MPT$ synthesis, 2-amino-4-hydroxy-6-hydroxymethyl-dihydropteridine diphosphokinase[76] reveals independent origins en route to LBCA and LACA (**Fig. 3A**). In methyl synthesis of the acetyl-CoA pathway, two reactions (bacteria) or three reactions (archaea) are catalyzed by structurally unrelated enzymes[67], not listed here as independent innovations because the reactions are not identical, differing by the pterin cofactor $H_4F$ vs $H_4MPT$, the syntheses of which themselves entail several unrelated enzymes[66]. Riboflavin synthase involved in flavin synthesis (**Fig. 3B**) also reveals independent origins in LACA and LBCA, representing a case in which the reaction requires no catalyst (**Table S4**), proceeding spontaneously at pH 7 upon boiling[77]. Increased temperature increases reaction rate for most biochemical reactions[59], an observation favoring a thermophilic origin of metabolism.

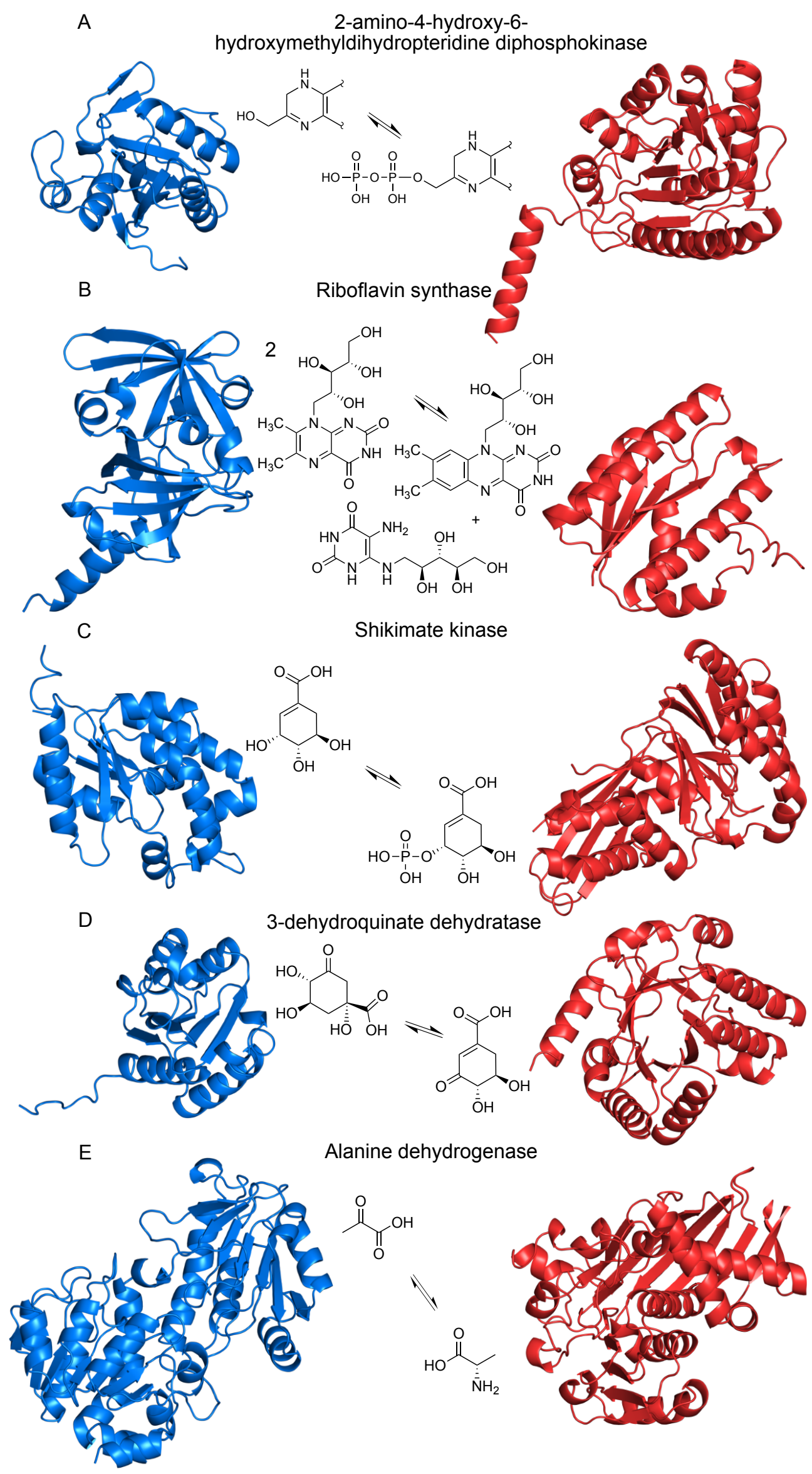

**Fig. 3. Protein families independently originated in the bacterial and archaeal domain.** Five cases of independent origins for structurally unrelated protein families, shown as the archaeal structure (red), the bacterial structure (blue), and a schematic of the catalyzed reaction (see main text). The pairs each generate a TM-score of ≤0.5 in the pairwise comparison using US-align (see **Methods**).

***Metals generated metabolism in serpentinizing systems***

The reactions surrounding the acetyl-CoA pathway and incomplete rTCA cycle take place in the presence of transition metals under serpentinizing hydrothermal conditions[26–28,31], but these pathways are not an exception in core metabolism. Among the reactions in our sample, 37 (10%) have been reported to proceed as in metabolism using the conditions and metal catalysts of hydrothermal vents **(Table S4)**. The metal catalysts (native and divalent transition metals) functionally replace enzymes and, in many cases, cofactors **(Table S4)** in the laboratory, for example NADH in carbonyl reductions to alcohols[78], or PLP in reductive amination of 2-oxoacids to amino acids[29,30]. For a further 106 reactions, the reaction type but not the exact reaction has been shown, for example reductive amination of carbonyls on alternative carbon backbones beyond the seven biological amino acid synthesis reactions demonstrated by Kaur *et al.*[29] **(Table S4)**. In 23 further reactions, the sequence from reactants to products proceeds under hydrothermal conditions, but the pathway in the laboratory involves a different number of chemical steps or a different reaction sequence. These abiotically catalyzed metabolic reactions mainly reside in carbon metabolism: the acetyl-CoA pathway[26,28,31,33], the incomplete reverse TCA cycle[27,29], glycolysis/gluconeogenesis[79] and the pentose phosphate pathway[80]. Yet many are involved in amino acid biosynthesis, supplying amino acids for synthesis of the first proteins.

While abiotic organic synthesis in modern serpentinizing systems is so far limited to the detection of formate[69,70,81], methane[82] and, more rarely, amino acids[83,84], serpentinizing systems are typically saturated with microbial life[69,70], often with an abundance of primary producers that use the acetyl-CoA pathway for carbon and energy metabolism, acetogens and methanogens[81,85], such that abiotic organics observed in effluent samples[69] are typically leftovers of microbial growth. One particularly important property of hydrothermal systems for the origin of metabolism is their microporous structure, providing the ability to concentrate organic reaction products at or near their site of synthesis, thereby enhancing further reaction[44,48]. The catalysts, reactants, reactions, and properties of serpentinizing hydrothermal vents have far-reaching congruence with the reactions of metabolism itself, suggesting that they represent the site, and the chemistry, of metabolic assembly.

***Phosphite is the source of phosphorylation during assembly***

Under the conditions of serpentinizing hydrothermal vents, 91% of metabolic reactions in our sample are exergonic in the biosynthetic direction (**Fig. S2, Table S5**), whereby 213 (59%) involve reactions of highly polarized, hence reactive carbonyl groups, and 134 reactions (37%) involve reactions of phosphate groups (**Table S3**). In modern metabolism, ATP is the currency of chemical energy that enables endergonic reactions to go forward[9,86]. In cells that use the acetyl-CoA pathway for carbon and energy metabolism, ATP is synthesized by the rotor stator ATP synthase, requiring ion-tight membranes and mechanisms that couple ion pumping to the exergonic reaction of $H_2$ with $CO_2$ (**Fig. S3, Table S6, supplementary text**). The source of high-energy phosphate bonds at metabolic origin, prior to the origin of membranes, is still discussed[23,87,88] because geochemically viable, continuous aqueous sources of phosphorylation have not been reported. Though reaction networks lacking phosphate in reactants and cofactors can be constructed using computers[25], the metabolic synthesis of most amino acids, all cofactors, and all nucleotides involve formation or cleavage of phosphate bonds—phosphate is a non-negotiable component at the origin of microbial metabolism. Acetyl phosphate, a possible primordial currency of high energy bonds[44], spontaneously forms from reactions of thioacetate and phosphate in water[89], though no continuous environmental source of thioacids is known. The growth of metabolism from a simple intermediate state (**Fig. 1B**, **Fig. 2**) prescribes a single environment for metabolic assembly[45,60], predicting the existence of a mechanism of phosphorylation that operates under serpentinizing conditions. In that search, microbial physiology informs.

Some bacteria can use phosphite ($H_2PO_3^-$; $P^{+3}$) as their sole source of phosphorus via the reaction $H_2PO_3^- + AMP + NAD^+ \rightarrow ADP + NADH$[38,39]. Phosphite occurs naturally in serpentinized rocks[12], is widely discussed as, but not demonstrated to be, a prebiotic source of organophosphate bonds[12,13], and genes for enzymes of phosphite oxidation are enriched in microbial communities of serpentinizing systems[90,91]. Phosphite is a very strong reductant, with a standard midpoint potential of –690 mV[39] vastly exceeding that of $H_2$ (–414 mV), but it is kinetically stable[12], meaning that it requires activation[38] in order to react. Because $Ni^0$ is effective in activating $CO_2$, $H_2$, $NH_3$, and organic molecules[26,28,29,31], we tested whether $Ni^0$ could also activate phosphite. Across a broad pH range (9–11), phosphite in water is readily converted to phosphate and $H_2$ by $Ni^0$, but not by $Fe^0$, $Co^0$, or magnetite, at roughly 75% yields overnight at 100°C (**Fig. 4A**).

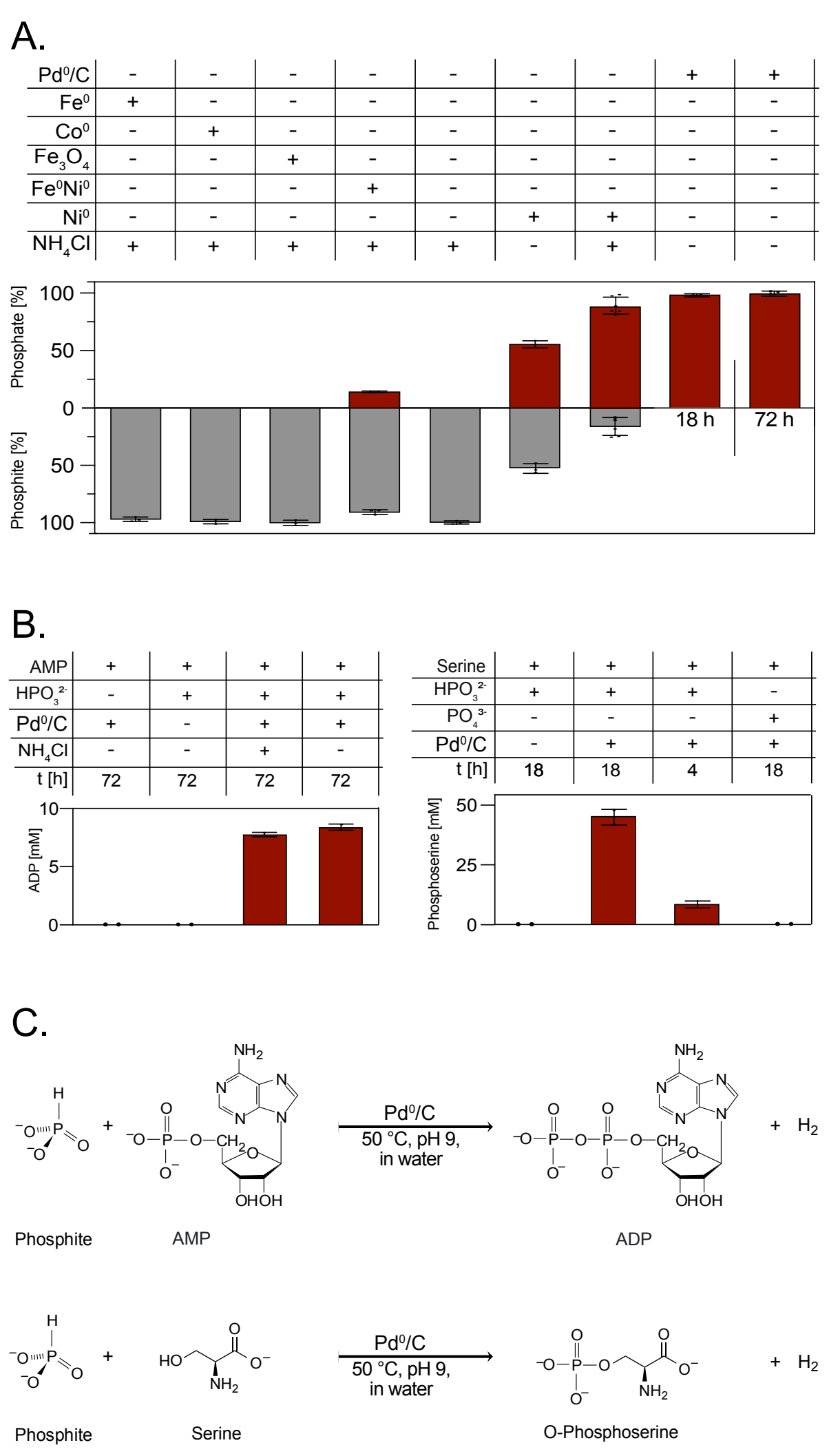

**Fig. 4. Aqueous, phosphite-dependent phosphorylation over metals. A**. Oxidation of phosphite (200 mM) to phosphate in water at pH 9, 100°C for 18 h or 72 h. Each reaction (triplicates) contained 1.5 mmol of Ni (in 1.5 mL) or 0.1 mmol of Pd/C (in 1.0 mL). $NH_4Cl$ concentration[24], when added, was 75 mM. **B**. Phosphite in water phosphorylates AMP (100 mM) and serine (100 mM) over Pd/C in water at pH 9 and 50°C (**Fig. S4**). Use of phosphate instead of phosphite yields no phosphorylation. **C**. Reaction of $HPO_3^{2-}$ with AMP and serine. Phosphite oxidation generates $H_2$. Turnover number for phosphite per Pd atom was 2, but not all Pd atoms are catalytically available and only a portion of phosphate-producing reactions phosphorylate AMP or serine.

If we react 200 mM phosphite with 100 mM AMP, replacing AMP-dependent phosphite dehydrogenase[38] with $Ni^0$ (1.5 mmol in 1.5 mL), we observe after 96 h at 50 °C and pH 7 phosphorylation of AMP in water to 1.05 μM adenosine diphosphate, partitioned into three isomers, the ATP precursor adenosine 5'-diphosphate, adenosine 3',5'-diphosphate and adenosine 2',5'-diphosphate (**Fig. S5**). The ~1 μM ADP yield demonstrates the surprising ability of $Ni^0$ to promote phosphorylation, but the yield is low, hence we tested its group 10 homolog, palladium. Like Ni, elemental Pd is an efficient catalyst of TCA cycle reactions and amino acid synthesis[29], and is naturally deposited in serpentinizing systems, often as a component of awaruite ($Ni_3Fe$)[36,37], which catalyzes $CO_2$ fixation[26,32]. In the presence of Pd (10% on carbon), phosphite in water phosphorylates the phosphate group of AMP at pH 9 to ADP with 8% yield in 72 h at 50 °C (**Fig. 4B**). The 8 mM concentration of ADP obtained with Pd/C is intermediate between the physiological concentrations of 560 μM ADP and 9.6 mM ATP in exponentially growing *E. coli* cells[92]. Phosphite over Pd/C also converts serine to phosphoserine, with 42% yield in 18 h at pH 9 (**Fig. 4C**). Other biological substrates are also phosphorylated under these hydrothermal vent conditions, including acetate to form acetyl phosphate[93]. In contrast to previous prebiotic phosphorylation protocols, no ammonium, cyanide, urea or other condensing agents (discussed in[23,24,88], are required for these metal-promoted phosphite-dependent phosphorylations, although 75 mM ammonium does not inhibit the reaction (**Fig. 4B**). Furthermore, $^{31}P$ NMR reveals that a small, but reproducible, portion of the ADP formed is further phosphorylated to ATP in the reaction over Pd/C (**Fig. S4**). These findings argue strongly in favor of a single, aqueous, phosphorylating environment for metabolic origin. For metabolic assembly, continuous, physiological-level phosphorylation in a hydrothermal system over hundreds of millennia is more life-like, and more biosynthetically useful than intermittent spikes of phosphides from meteoritic delivery[23], because metabolic assembly required protracted geological time, continuous energy release[94], continuous phosphorylation and a continuous environment for enzymatic invention (**Fig. 2**).

Native nickel emerges as a remarkably multifunctional, broad-spectrum catalyst for the incorporation of $H_2$, $CO_2$, $NH_3$, and now (though less effective than Pd) phosphorus, into nascent and primordial metabolism. As $Ni^{2+}$, nickel is present in the active site of ancient enzymes of acetogens and methanogens: hydrogenases[46], CODH and acetyl-CoA synthase of the acetyl-CoA pathway[61,62], as well as methyl-CoM reductase in methanogenesis[95]. As $Ni^0$ it functionally replaces the entire acetyl-CoA pathway, catalyzing nonenzymatic $CO_2$ reduction to pyruvate[26,28,33,57], TCA cycle reactions[29], reductive aminations in amino acid syntheses[29], and promotes the phosphite-dependent phosphorylation of phosphate and hydroxyl groups in AMP (**Fig. S5)**. The phosphorylation reaction is thermodynamically favorable because the free energy of phosphite oxidation to phosphate, $HPO_3^{2-} + H_2O \rightarrow HPO_4^{2-} + H_2(g)$ with $\Delta G_0 = -53$ kJ·mol$^{-1}$ at pH 7[23], is sufficient to generate a phosphoanhydride bond in ADP[38,94], without the participation of external energy sources. Phosphite activation involves hydride removal from phosphite. Though neither $Ni^0$ nor $Pd^0$ are oxidants, both are good hydride acceptors and excellent catalysts of reversible $H_2$ synthesis[96]. These findings show that metals naturally deposited in serpentinizing systems ($Ni^0$ and $Pd^0$) promote the oxidation of phosphite, which is formed during serpentinization[12], to phosphate in such a manner as to phosphorylate phosphate- and hydroxyl-groups at physiological temperature and pH in water. Native metals of serpentinizing vents transform $H_2$, $CO_2$, $NH_3$ and $HPO_3^{2-}$ into phosphorylated organic compounds in reactions similar or identical to reactions of metabolism.

### *Cofactors replace environmental catalysts*

Cofactor synthesis is conspicuously incomplete in LUCA's metabolism **(Fig. 2)**. Though traditionally thought to precede enzymes in biochemical evolution, and while clearly effective in promoting reactions without enzymes[14,42,45], cofactors do not readily arise spontaneously in aqueous prebiotic-type reactions[14]. They have complicated biosynthetic pathways in metabolism[97], their chemical synthesis in the laboratory has to be meticulously orchestrated[54], and in metabolism they are often required for their own enzymatic synthesis[14,44,98]. At face value, these observations speak against the notion that cofactors are either older than enzymes[42], relics from an RNA world[99]. In core metabolism, most cofactors do not act as catalysts, rather they participate as reactants that transfer mass, chemical moieties, to or from substrates[42]. Moreover, in the presence of native metals, recent reports show that several cofactors even accept donor moieties directly from metal surfaces, uncovering insights into the primordial assembly of metabolism.

For example, $H_2$ in the presence of $Ni^0$, $Fe^0$ and $Co^0$ converts $NAD^+$ to NADH in water[46,100] and $Ni^0/H_2$ can replace NAD(P)H as a reductant for numerous redox reactions of core metabolism that involve hydride transfer[29,30]. The synthesis of low potential reduced ferredoxin, required in five core metabolic reactions, involves a multi-enzyme system of flavin-based electron bifurcation and one-electron transfers in cells[86,101], but ferredoxin can be reduced with single electrons stemming from $H_2$ and $Fe^0$ in water[102]. Pyridoxal phosphate (PLP)[47] performs roughly 20 enzymatic aminotransferase reactions in metabolism (**Table S3**), is replaced by $Ni^0$ or $H_2/Ni^0$ in the presence of $NH_3$[29] (**Fig. S6**) and is reductively aminated to pyridoxamine with $Ni^0$, $H_2$ and $NH_3$ in water under hydrothermal conditions[29]. In the reactions of $CO_2$ and $H_2$ over $Ni^0$, $Fe^0$ or $Ni_3Fe$ that replace roughly 10 enzymes and 10 cofactors of the acetyl-CoA pathway[57], the molybdenum cofactor, MoCo, required for formate synthesis from $CO_2$[63] is replaced by Ni/$H_2$, as is thiamine, required for pyruvate synthesis in the pathway[28,103]. The same is true for the C1 carriers tetrahydrofolate ($H_4F$) and tetrahydromethanopterin ($H_4MPT$), which are required for methyl synthesis from $H_2$ and $CO_2$[26] in bacteria and archaea[66], and the cobamide cofactor (Cob)[66] required for metal-to-metal methyl transfer in the acetyl-CoA pathway[104]. In $Ni_3Fe$-dependent methane synthesis[26], the alloy replaces the function of the Ni-containing tetrapyrrole $F_{430}$ of methanogenesis[95].

Though this list of metal-catalyzed metabolic reactions is short, a general principle clearly emerges: native metals activate and transfer $H_2$-, $CO_2$- and $NH_3$-derived moieties to substrates, generating biochemical products of metabolism and thereby functionally substituting for essential cofactors of core metabolism under hydrothermal conditions. Furthermore, in cases experimentally studied so far, the cofactors interact specifically with metal surfaces to obtain the activated moiety that they transfer to the substrate: hydride in the case of NADH[46,100], single electrons in the case of ferredoxin[102], amino groups in the case of pyridoxal[30], and phosphate in the case of ADP (**Fig. 4B**). Native metals are clearly simpler and older than cofactors, and are functionally equivalent to cofactors for group transfers in cases studied so far.

***Cofactors replace metals, but are synthesized by enzymes***

Biosynthetic pathways should hold clues about the order in which catalysts arose[14,41,45]. To address this issue systematically across core metabolism, we serially ordered the appearance of all reactions during metabolic assembly, starting at the acetyl-CoA pathway and leveraging the principle that the appearance of a reaction requires the existence of its reactants as products of earlier reactions (see **Methods** and **supplementary text**). For reactions that require cofactors as reactants, their metal-catalyzed moiety-donating equivalents were coded as being provided by the environment until the cofactor was synthesized, at which point it replaces the metallic forerunner. Because a metabolic network can grow by one reaction at several nodes simultaneously, reactions of equal age are grouped into the same time unit, or stratum, whereby the ensuing order and occupancy of strata are unique. The ordered list of enzymatic reactions in metabolism is given in **Table S7**. The order in which products arise is of interest, in particular the relative timing of cofactor synthesis. As expected, amino acids arise first, followed by nucleotides, consistent with their inferred synergistic interactions at origins[105], with cofactors emerging last (**Fig. 5A**). Altogether unexpected, however, was our finding that cofactors not only arise late, all cofactors are required in reactions that precede their synthesis in metabolism (**Fig. 5B**). For 9 out of 17 cofactors, the cofactor is synthesized in the last reaction of core metabolism in which it occurs. In total, cofactors are required for 209 reactions that precede their synthesis as opposed to 90 reactions that follow their synthesis. Moreover, 12 out of 17 cofactors, including the last cofactor synthesized—coenzyme A—are required in the first 7 reactions at the starting point of metabolism: the bacterial and archaeal versions of the acetyl-CoA pathway (**Fig. 5B**). This organization of metabolism seems staggeringly illogical, it demands a simple explanation.

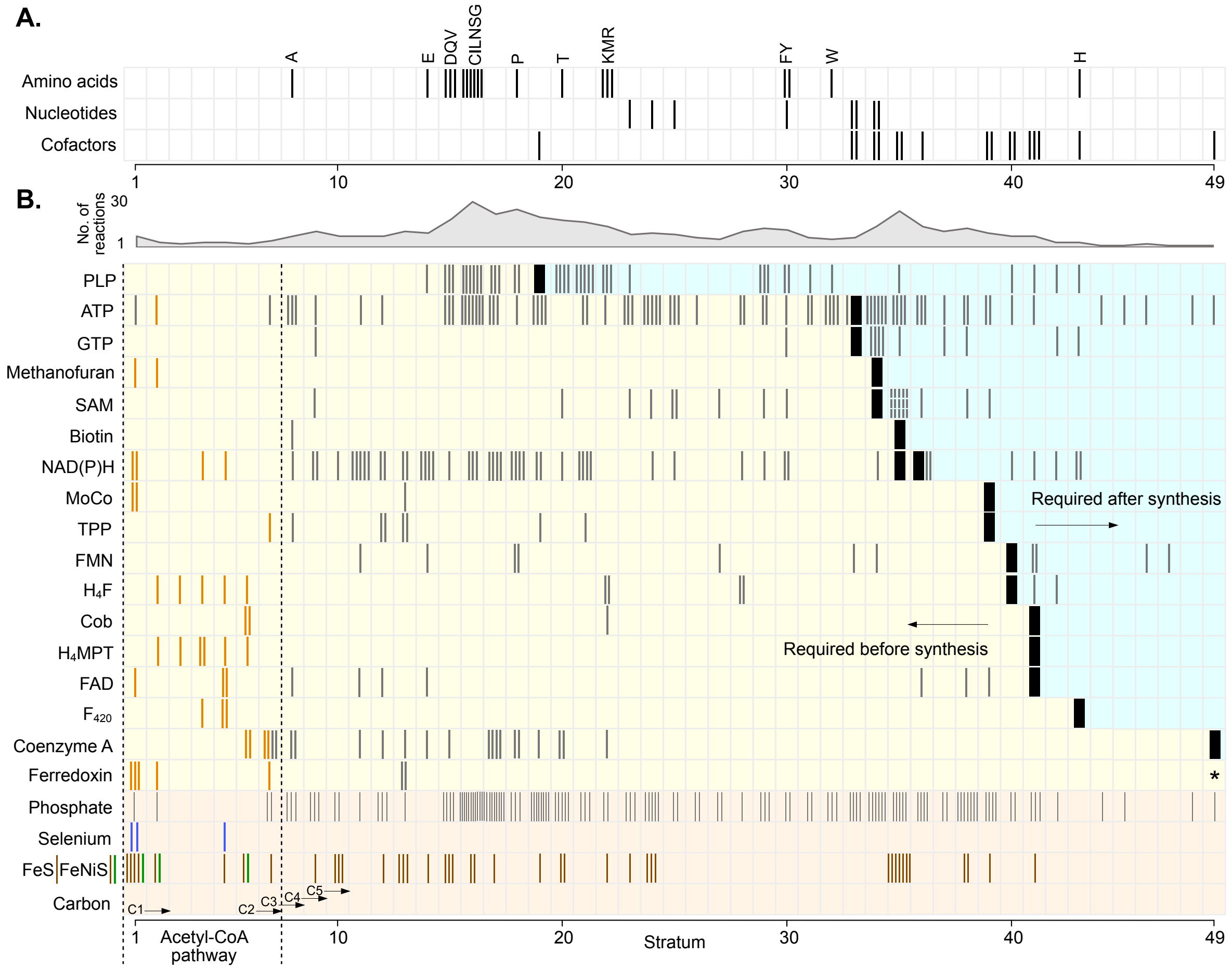

**Fig. 5. Serial order of metabolic products.** Reactions were ordered by their appearance in metabolism (see **Methods**). **A.** The appearance of amino acids, nucleotides (from left to right UTP, CTP, dCTP, dTTP, ATP/GTP, dATP/dGTP), and cofactors (order given in panel **B**). See also **Fig. 6**. **B.** The number of reactions appearing in each stratum (1–49) is plotted. The matrix shows the order of appearance of reactions that require the given cofactor (narrow vertical lines) and reactions that synthesize the cofactor (solid vertical lines). Strata before the synthesis of the cofactor are shown in yellow, while strata following the synthesis of the cofactor are in blue. Vertical lines in sepia indicate reactions of the acetyl-CoA pathway, which are restricted to strata 1–7 (dotted lines). Ferredoxin is a protein with FeS clusters synthesized by ribosomes, not metabolism. Coenzyme B, coenzyme M and $F_{430}$ are not included because they are not involved in biosynthetic metabolism. Five out of the six ATP-dependent reactions at stratum 34 correspond to donations of adenosyl or AMP residues. The order of amino acids is the order of their appearance in metabolism catalyzed by enzymes, recalling that the first enzymes were synthesized from non-enzymatically formed amino acids. References to metal-catalyzed reactions that serve as evolutionary precursors to cofactor-dependent reactions are given in **Fig. S7**. Lower

panel: Reactions involving phosphate bonds, reactions catalyzed by selenoproteins, reactions catalyzed by proteins with FeS or FeNiS clusters, first appearance of C1–C5 carbon backbones.

As one possibility, there could have existed more ancient alternative syntheses[15] for each cofactor, using starting compounds not found in biology, for example in the case of cyanide-dependent synthesis of coenzyme A (CoA)[54], that might have provided organic placeholders for the biosynthesized cofactor at 23 different reactions up until its synthesis via enzymatic metabolism at stratum 49 **(Fig. 5B)**. Once CoA became synthesized by $CO_2$-dependent metabolism, the ancestral cyanide-derived CoA would no longer have been needed, although a cyanide-dependent synthesis cannot have coexisted with metabolic $CO_2$-dependent CoA synthesis, because the acetyl-CoA pathway is inhibited by cyanide[58,106]. Although such possibilities are entertained[15,54], consider that in laboratory reactions, transition metals can catalyze thioester bond formation and cleavage using simple organic thiols, such as alkyl sulfides, which exert the salient function of CoA[4,107], or catalyze the entire acetyl-CoA pathway[26,28,31,57]. Why should metabolism have required a complicated, highly orchestrated, highly specific, abiotic synthesis of many (identical) molecules of CoA continuously over geological time if ubiquitous environmental thiols and metals can fulfill the function? The far simpler, metal-catalyzed thiol dependent reaction can, in principle, serve instead of the cofactor until the latter is synthesized by metabolism.

The same principle comes to bear for all other cofactors. For example, NAD(P)H, is required in 47 reactions preceding its synthesis (**Fig. 5B**), but can be replaced[46,100] by $H_2/Ni^0$ in 29 reactions of metabolism (**Table S4**), including reductive aminations[29]; NADH ($C_{21}H_{27}N_7O_{14}P_2$) exerts the function of irreducibly simple compounds, $H_2$ and $Ni^0$, present in the environment. Tetrahydrofolate ($H_4F$; $C_{19}H_{23}N_7O_6$), is required for one-carbon transfers in five reactions of the acetyl-CoA pathway (**Fig. 5B**), but its methyl synthesis and transfer function is replaced by $H_2$ over $Ni^0$ or $Ni_3Fe$ during aqueous pyruvate formation from $CO_2$[26,28,31]. Metabolic assembly did not require abiotic $H_4F$ for pyruvate formation, $H_2$ over native metals was sufficient. PLP is required in 15 reactions prior to its synthesis (**Fig. 5B**), but $H_2/Ni^0$ or $Ni^0$ alone in the presence of $NH_3$ can replace the cofactor[29] (**Fig. S6**) in 21 laboratory metabolic reactions (**Table S4**) and reductively aminate it to the active pyridoxamine form[30]. As a final example, the molybdenum cofactor MoCo ($C_{10}H_{12}MoN_5O_8PS_2$), or its tungsten homologue Wco ($C_{20}H_{22}WN_{10}O_{16}P_2S_4$), are required for formate synthesis from $CO_2$ in the first step of the acetyl-CoA pathway in bacteria and archaea, but are also replaced by $H_2$ over $Ni^0$, $Co^0$, $Fe^0$, or $Ni_3Fe$[26,28], all of which are naturally deposited in serpentinizing hydrothermal vents, which also provide $H_2$ as reductant[34]. Metabolic assembly did not have to await the synthesis of an organic precursor to MoCo/Wco in order to make formate, which is still synthesized today in serpentinizing hydrothermal vents[69,70] , nor did metabolism have to await the synthesis of NADH, $F_{420}$, $FADH_2$, or reduced ferredoxin for reductions, thiamine pyrophosphate (TPP) for pyruvate synthesis from $H_2$ and $CO_2$, or ATP as a source of phosphorylation (**Fig. 5**), because native metals and hydrothermal chemical constituents alone perform the reactions, and phosphite-dependent phosphorylations conserve energy in the form of high-energy bonds. These findings are at odds the view that cofactors are remnants of a hypothetical RNA world that existed before enzymes[99]. Rather, cofactors appear as organic replacements for broad spectrum metal catalysts; enzymatically synthesized, they liberated primordial metabolism from its dependence upon solid state metals that fostered its assembly.

***Limitations of the study***

Identification of archaeal functions is a limitation in current data. Many archaeal enzymatic functions in core metabolism are not yet known[75,97]. Our analysis was restricted to enzymes of metabolism, hence limited to clusters having annotations of known biochemical function in at least one archaeal lineage. With expanded knowledge of functionally characterized enzymes of biosynthetic pathways in archaea, many more archaeal and archaeal specific enzymes[75,97] will likely be uncovered. This results in an apparent excess of bacterial specific inventions. In addition, archaeal lineages are less well sampled than bacterial, compounding the same sampling related problem. The list of experimentally demonstrated, environmentally catalyzed reactions at origins is still incomplete. Metals will likely not replace all of these enzymes, field catalysts[108,109] will likely be required in many instances.

At current count, 46% of core metabolic reactions and geochemical analogues have been shown to take place without enzymes (and cofactors) under aqueous conditions of hydrothermal vents and transition metal catalysis (**Table S4**), indicated as filled nodes in **Fig. 6**. The early synthesis of the Ni-containing tetrapyrrole $F_{430}$ required in the last step of methanogenesis may seem surprising given the late synthesis of Co-containing cobamides required in the acetyl CoA pathway (**Fig. 6**). The difference stems from the synthesis of the lower ligand in cobalamin[67], which is lacking in $F_{430}$. Cofactors that contain an AMP moiety such as NAD, FAD, SAM (S-adenosyl methionine) or CoA, arise subsequent to the synthesis of ATP, underscoring the requirement for an environmental source of phosphorylation at metabolic origin – $HPO_3^{2-}$ with Pd/C is such a source. A critic might contend that the abundance of Pd in the crust, $1.5 \cdot 10^{-8}$, is too low for relevance at origins, but physiology informs otherwise. Selenium is just as rare as Pd, $5 \cdot 10^{-8}$, but despite crustal paucity, Se became incorporated into the primordial genetic code as selenocysteine. Moreover, three reactions of the acetyl-CoA pathway are catalyzed by selenoproteins[110] (**Fig. 5B**), indicating that elements as rare as Se were continuously available in the environment where metabolism and the genetic code arose; $Pd^0$ is enriched in serpentinizing environments[36,37].

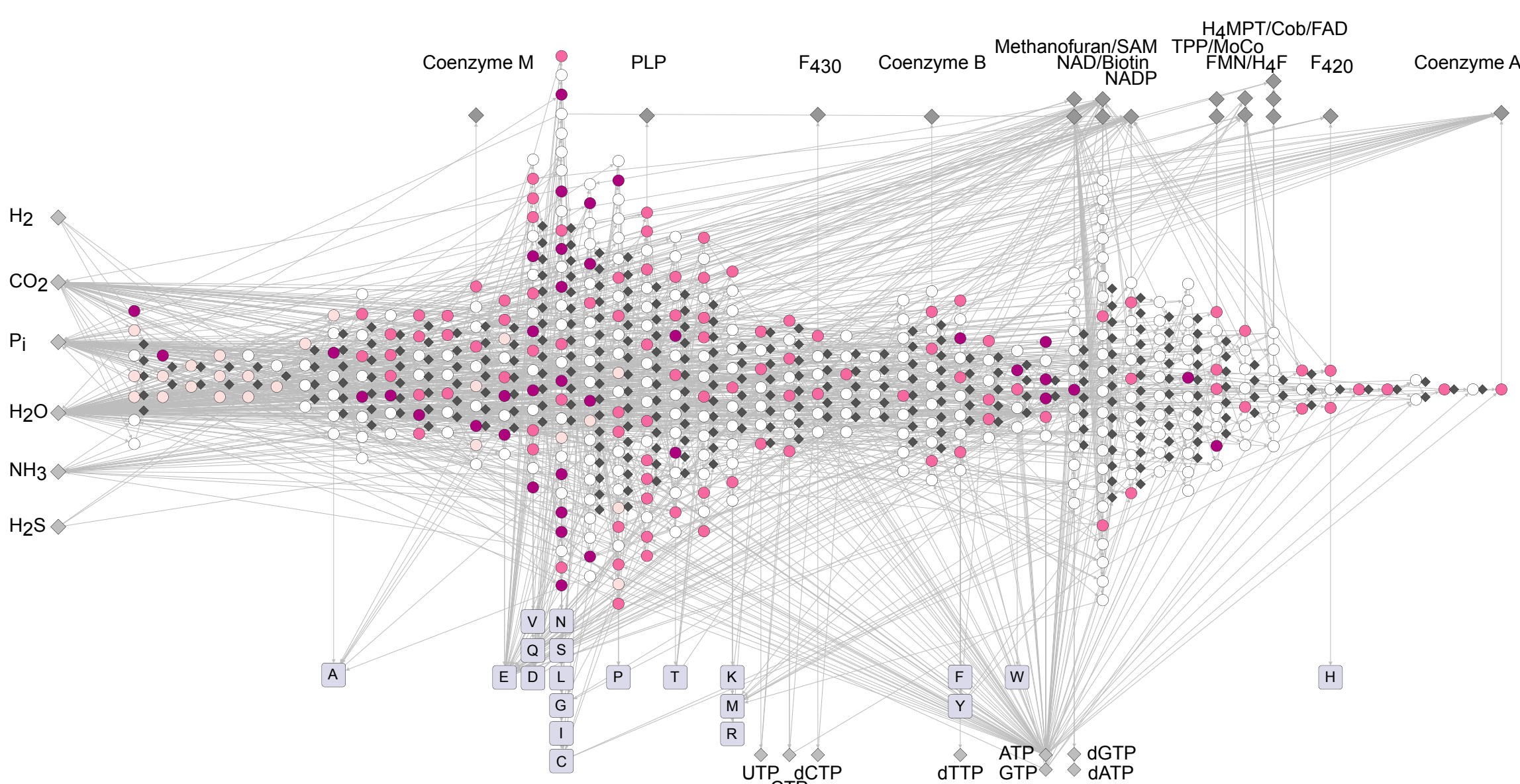

**Fig. 6: Structure of metabolic assembly**. Network of reactions stratified as in **Fig. 5**. Inorganic substrates are shown at left. Network end-products are shown at top (cofactors) and at bottom (nucleotides and amino acids indicated with one-letter codes). Diamond-shaped nodes indicate chemical compounds, circular nodes indicate reactions. Reactions shaded according to reactions and all corresponding references in **Table S4**: (i) (dark) the reaction has been experimentally shown to be catalyzed by native metals or is uncatalyzed, (ii) (light shading) part of a pathway or reaction sequence from biological substrates to products, is catalyzed by native metals in one-pot reactions, (iii) (intermediate shading) the reaction type (for example transamination) has been shown to be catalyzed by metals in water or (iv) the nonenzymatic reaction is not yet experimentally demonstrated (open circles).

If the 17 cofactors summarized in **Fig. 5**, and required in **Fig. 6**, all of which including ATP have metal-dependent functional precursors, were not needed at the geochemical onset of metabolism, why were they needed at all, and why in the terminal phases of metabolic assembly via enzymes? The answer, we propose, is that the cofactors were required as a chemical replacement for the native metals that enabled metabolic assembly, hence cofactor synthesis from the outset, such that metabolism could become independent of solid-state catalysts deposited in the Earth's crust. By severing the connection between metabolism and environmental catalysts, enzymatically synthesized cofactors rendered metabolism autocatalytic—generating its required catalysts by itself. The implication is that, in the extreme, the highly conserved organic cofactors of metabolism did not exist in their present form prior to the origin of ribosomal protein synthesis, which does not directly require cofactors other than GTP[51]. This would not preclude the existence of structurally simpler but functionally equivalent precursors cofactors which later became replaced[45,111], as discussed for replacement of PLP by TPP in some reactions[112].

The gradual assembly of metabolism from environmental reactions to reactions present in LUCA and the independent growth of metabolism in the lineages leading to LBCA and LACA is outlined in **Fig. 7**. As metabolism underwent assembly, enzymes and cofactors gradually replaced metals and reaction rates (an unwieldy variable for environmental metals) came under the control of natural selection. Enzymes and cofactors provided the substrate specificity[113] required for metabolism to operate as a unit and fostered its hallmark: all reactions proceed at roughly the same rate[59] so that carbon and energy metabolism could form a tightly connected, stoichiometrically balanced[86] reaction. Given that native metals ($Co^0$, $Ni^0$, $Fe^0$, $Pd^0$) are excellent prebiotic catalysts, performing the reactions of 17 different cofactors (**Fig. S7**), why did they not become incorporated into enzymes during metabolism? Their specificity is too low, they catalyze too many different reactions well[114] (**Fig. S7**), making them a poor choice for prosthetic groups of enzymes.

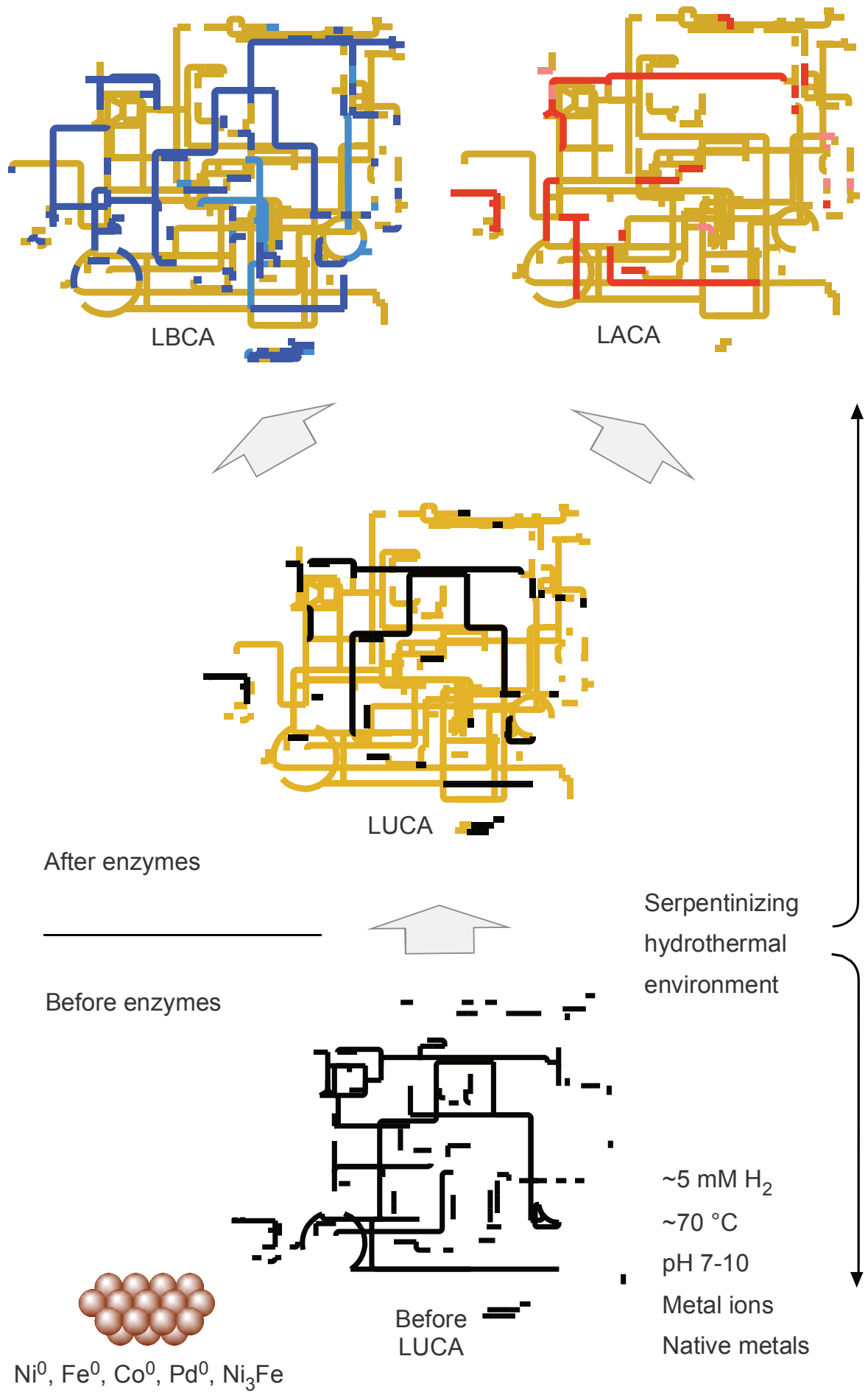

**Fig. 7. Assembly of metabolism at a phosphorylating hydrothermal vent.** Maps prepared with iPath (see **Methods**) show reactions catalyzed by enzymes present in LUCA (gold), lineage-specific additions in LBCA (blue) and in LACA (red). Lighter shaded reactions in LACA and LBCA indicate enzymes in grey areas of **Fig. 1B**. Black lines denote reactions that are either (i) present as enzymatic in LUCA, LBCA or LACA and can be catalyzed by native or divalent transition metals (see **Fig. S7 and Table S4**) before enzymes (lower panel), or that are (ii) present as enzymatic in either LBCA or LACA, but not in LUCA, and can be catalyzed by native or divalent transition metals in LUCA. Catalysis is provided by enzymes and native metals (symbolized at lower left), which catalyze one-electron transfer, hydride transfer, $H_2$ activation, $CO_2$ reduction, C1 transfer, formate synthesis, acetate synthesis, pyruvate synthesis, aldol condensation, decarboxylation, amino acid synthesis and amination (see **Fig. S7**), in addition phosphite activation and phosphorylation (**Fig. 4**).

Many theories for metabolic origin posit that FeS minerals in the environment served as catalysts prior to the origin of enzymes[3,44,45], yet FeS clusters of modern proteins serve mainly as mediators of electron transfer[63]. The present findings indicate that the essential catalysis of origins was exerted not by metal sulfides, but by native metals deposited in serpentinizing hydrothermal systems[34–37]. During metabolic assembly, they were replaced by enzymes and cofactors, the latter serving as soluble replacements for native metals in the rocks-and-water environment where metabolism and cofactors arose: a $H_2$-producing and phosphorylating hydrothermal vent.

## Materials and Methods

### *Data acquisition*

Prokaryotic whole genome protein sequences were downloaded from the RefSeq database (release 223, May 2024). The data was filtered for the largest genome per species for archaea and the largest genome per family for bacteria. Subsequently, the list of acetogens without cytochromes from Rosenbaum and Müller[115] was checked for organisms (exact species and strain) not present in the filtered dataset, but whose genomes were present in the downloaded complete RefSeq dataset. These were added to our filtered dataset to account for underrepresentation. Genomes with a size larger than 1000 proteins were retained (to exclude parasites). This yielded a balanced dataset of 401 archaeal and 552 bacterial genomes (see **Data availability**). In-house Python scripts, in some cases modified and debugged with ChatGPT 4o and Deepseek V3 and R1, were used for data processing throughout this work.

Reactions and E.C. numbers for core metabolic reactions that generate nucleotides, cofactors and amino acids from $CO_2$, $H_2$, $NH_3$, $H_2S$ and $P_i$ were obtained from Wimmer *et al.*[18]. E.C. numbers were updated according to KEGG (release 112.0, November 2024). E.C. numbers for enzymes involved in archaeal and bacterial glycolysis and gluconeogenesis were obtained from Bräsen *et al.*[116], the reaction IDs were retrieved from KEGG, and this reaction set was added to the list of core metabolic reactions. The archaeal $H_4$MPT-dependent variant of the Wood-Ljungdahl pathway methyl branch (from KEGG module M00567) was also added, resulting in an updated list of 424 core metabolic reactions (**Table S8**). Reactions added in this study were polarized in the biosynthetic direction, except for irreversible glycolytic reactions (based on Bräsen *et al.*[116]).

*Clustering and annotation of protein families in the biosynthetic core*

Protein sequences from 953 prokaryotic genomes were clustered into protein families as described in Brueckner and Martin[117], with DIAMOND-blastp[118] used instead of BLAST. The results were filtered with an e-value cutoff of $\leq 10^{-5}$ and a local identity cutoff of ≥25%. Hits with a global identity cutoff ≥25% were used for Markov Chain clustering. Out of the resulting 245,258 clusters (see **Data availability**), only those containing 4 or more sequences were retained, resulting in 103,482 MCL clusters.

To annotate the clusters, bacterial and archaeal sequences corresponding to the E.C. numbers of 424 core metabolic reactions (**Table S8**) were retrieved from SwissProt (release 2024_04) using the API. MCL clusters were aligned using DIAMOND blastp against the SwissProt dataset, and the results were filtered for ≥40% sequence identity and an e-value $\leq 10^{-5}$. After filtering, each sequence in the clusters was annotated with the E.C. of its best hit, if there was one. The entire cluster was assigned an E.C. number when a minimum of 50% of its sequences were annotated with that E.C.

For all E.C. numbers corresponding to the 424 core metabolic reactions that did not annotate a cluster, the corresponding prokaryotic sequences from UniProt[119] were downloaded through the API filtering for levels 1, 2 and 3 of protein existence evidence (to exclude predicted and uncertain proteins). Overall, database sequences for 369 (out of 389) E.C. numbers were obtained. The annotation workflow was repeated as described above, now using the sequences downloaded from both SwissProt (release 2024_04) and UniProt (release 2024_06) as the target database. This yielded 1473 MCL clusters annotated with 335 E.C. numbers corresponding to 361 metabolic core reactions. The enzyme name was obtained from KEGG based on the E.C.

*Merging of protein families in the biosynthetic core*

MCL clusters that correspond to the same reaction (or at least one of the reactions they are mapped to is the same) were examined to determine if they could be combined. In the first step, sequence similarity-based merging was carried out. Clusters mapping to the same reaction were categorized into small (<50 members) and large (≥50 members) clusters. An all-vs-all DIAMOND blastp was performed to determine how many members of the small clusters matched the large ones and how many members of large clusters matched other large clusters. The cutoff was set to ≥25% sequence identity and an e-value of $\leq 10^{-5}$. If more than 50% of sequences in a small or large cluster was above threshold in the alignment to a large cluster, the clusters were merged (and with them, all reactions they map to) (**Fig. S1, Table S9**).

In the second step, structural similarity was evaluated for clusters that correspond to the same reaction but could not be joined based on sequence similarity. A representative sequence for these clusters was chosen based on its highest average local identity to all other cluster members out of the sequences with the highest number of BLAST hits in the cluster. Structures for the representative sequences were modelled with AlphaFold 2.2.0[120], ColabFold[121] or retrieved from the AlphaFold database[122]. Representative structures of clusters mapped to the same reaction(s) were aligned using US-align[123]. TM-scores normalized by the length of both sequences in the comparison were obtained, and if at least one of the two TM-scores was ≥0.5, the structures were considered to share a common fold[124] (**Table S9**), and these clusters were merged (**Fig. S1**, see **Data availability**).

Clusters mapped to the same reaction could correspond to subunits of the same enzyme complex, so manual verification of subunit information was conducted based on the sequences used for the annotation of each cluster.

*Assignment criteria for biosynthetic core protein families and ΔG calculation*

The taxonomic distribution of each protein family (cluster) was calculated as the percentage of genomes in each higher taxon (archaeal orders and bacterial classes) that contain a member of that protein family. The artificial taxa "Other Bacteria" and "Other Archaea" were introduced for genomes from very small higher taxa containing less than 4 genomes, or genomes with no information on taxonomic order/class. Clusters present in <10% of archaeal and <10% of bacterial genomes were classified as rare protein families that provide limited information and were not considered further (listed in **Table S2**). The distribution table with the remaining clusters was sorted by the quotient 15*(sum of bacterial distribution percentages in the cluster)/26*(sum of archaeal distribution percentages in the cluster), where 15 is the number of archaeal and 26 is the number of bacterial higher taxa. Zero denominators were replaced with 1. The table was sorted left-to-right based on increasing gene frequency per column (taxon) in Bacteria, and decreasing gene frequency per column (taxon) in archaea.

The following gene retention criteria were applied: if the gene (cluster) was retained in ≥50% of genomes of ≥4 bacterial, but not archaeal higher taxa, it was classified as LBCA. If, on the other hand, it was retained in ≥50% of genomes of ≥4 archaeal, but not bacterial higher taxa, it was classified as LACA. A cluster was categorized as tracing to LUCA if it occurs in ≥50% of genomes of ≥4 taxa in both bacteria and archaea. The LBCA and LACA clusters were resorted based on the total percentage of bacterial and archaeal members, respectively. These conservative criteria of gene retention resulted in two uncertain regions that include clusters that are predominantly bacterial, but do not satisfy the LBCA criteria (the grey LBCA region) and those that are predominantly archaeal, but do not satisfy the LACA criteria (the grey LACA region) (**Table S2**). Clusters that did not satisfy the gene retention criteria were manually moved to the grey categories, resulting in the distribution in **Table S2** and **Fig. 1B**.

Cluster 692, corresponding to alanine dehydrogenase, is a borderline case because it is present in ≥50% of genomes of 4 bacterial classes, half of which contain a small number of genomes: 4 in Myxococcia and 6 in Thermoleophilia, with the protein family present in 50% of both. This distribution and previous work reporting the unrelatedness of the archaeal and bacterial enzyme[125], classify cluster 692 as independently arisen in **Fig. 3**.

The taxonomic distribution data were visualized in a matrix (**Fig. 1B, Fig. 2, Fig. S1**) with the Python packages Matplotlib, Pandas and Seaborn. Protein structures in **Fig. 3** were rendered with Pymol (The PyMol Molecular Graphics System, Version 2.5.4., Schrödinger, LLC). Metabolic maps for **Fig. 7** were prepared with iPath[126]. Reactions corresponding to protein families assigned to LUCA, LBCA and LACA **(Table S2)** and reactions that could have an environmental precursor (**Table S4**) were visualized. A caveat is that the tool could not display some KEGG reactions in the maps. Values for ΔG under different conditions (**Fig. S2, Table S5**) were calculated as described in Wimmer *et al.*[18]. Literature references are listed for manually curated information (**Table S3**, **Table S4,** enzyme references and enzyme promiscuity assignment in **Table S2)**.

*Gene distribution and structure visualization of ribosomal proteins*

Protein families were not annotated with ribosomal proteins sequences, as they are not part of the autotrophic biosynthetic core. For ribosomal proteins (**Fig. 1A; Table S1**), known to be short and biased in composition[127], all SwissProt prokaryotic ribosomal protein sequences of protein existence level 1 identified by protein name (**Table S1**) were retrieved. These sequences were used as queries in DIAMOND blastp alignments against the balanced dataset of 401 archaeal and 552 bacterial genomes (see *Data acquisition*). Hits were considered homologs if they exhibited a local sequence identity ≥25% and an e-value of $\leq 10^{-10}$.

Ribosomes from *E. coli* and *T. kodakarensis* are taken as models for a bacterial and an archaeal ribosome (**Fig. 1A**), therefore ribosomal proteins from these ribosomes were extracted from the BLAST hits file filtering by ribosomal protein names (**Table S1**), and a taxonomic distribution matrix was generated. Universal and domain-specific ribosomal proteins were identified based on the literature[128–130]. When BLAST searches were not sensitive enough to detect homologies known from the structural literature, rows were manually joined to reflect known evolutionary relationships. Figures of protein structures (PDB ID: 7k00[131]; PDB ID: 6SKF[130]) were rendered with Pymol (The PyMol Molecular Graphics System, Version 2.5.4., Schrödinger, LLC). Domain-specific components were colored according to[128–130].

*Stratification of reactions and compounds*

An upgraded version of the software CatReNet (now v. 0.10.1)[132] used for the analysis is available at https://software-ab.cs.uni-tuebingen.de/download/catrenet/welcome.html. The new release was designed to include a stratification algorithm based on the following formal framework (which extends concepts from Steel *et al.*[133]).

Let $X$ be any set of molecule types that can arise by starting from a food set $F$ of basic molecule types and repeatedly applying reactions from some given reaction set $R$. The number of possible orderings of the reactions that will generate all the molecule types in $X$ could be astronomical; however, there is a well-defined underlying partial order on (i) the reactions and (ii) the elements of $X$. These two orders reveal how early any given reaction could possibly have occurred, and any given molecule type could have first appeared. Moreover, there is a fast way to compute these partial orders, and they are uniquely determined by the system (so multiple runs of the algorithm are unnecessary). We refer to this process as *stratifying* the sets $R$ and $X$.

Formally, given reaction system with food set $(X, R, F)$ where $R$ is $F$–generated, let $R_0 = \{r \in R: \rho(r) \subseteq F\}$, where $\rho(r)$ is the set of reactants of $r$. For $i \geq 1$, let:

$$R_i = \left\{ r \in R: \rho(r) \subseteq F \cup \pi\left( \bigcup_{0 \leq j < i} R_j \right) \right\}$$

The sets $R_i$ are 'nested' (i.e. $R_i \subseteq R_{i+1}$ for all $i$), and since $R$ is $F$–generated, $R_i = R$ for some value of $i$. For each. $r \in R$, we can then define an associated ranking $\lambda(r)$ to be the smallest value of $j$ for which $r \in R_j$.

We can also stratify $X$ for any $F$–generated set $R$, as follows. Let $X_0 = F$, and for each $i \geq 1$, let:

$$X_i = X_{i-1} \cup \left( \bigcup_{r \in R: \rho(r) \subseteq X_{i-1}} \pi(r) \right).$$

Thus $X_i$ is $X_{i-1}$ together with all the products of all the reactions in $R$ that have all their reactants in $X_{i-1}$. The resulting collection of subsets of $X$ thus forms a nested increasing sequence of molecule types: $F = X_0 \subset \cdots \subset X_K = X_{K+1}$. Next, define $\lambda': \pi(R) \rightarrow \{1,2,\ldots\}$ by:

$$\lambda'(x) = \min\{i: x \in X_i\}.$$

In this way, the elements of $\pi(R)$ are partitioned by their $\lambda$ values, and $\lambda'(x)$ is a measure of how early element $x$ can be generated (further technical details are provided in the **supplementary text**).

In order to determine the order of emergence of core metabolic reactions, the reaction set had to be adapted to form a continuous network (for a list of excluded and included reactions see **Table S7**). Co-substrates such as NADH or reduced ferredoxin were included in the food set (for a full list see **Table S7**), as previous work has shown that their functions could be replaced by mineral surfaces at the origin of metabolism (see main text and **Fig. S7**). Compound recoding was carried out in those cases where it was necessary to distinguish between the environmental version (the function of the cofactor is carried out by mineral surfaces) and the synthesized, organic version of a cofactor or a product generated from it (**Table S7**). The point of synthesis of a cofactor marks the start of its employment in the network in place of the environmental precursor. All reactions that require a cofactor and occur after its synthesis naturally employ the organic version, even if it is not specified in the recoding. The complete stratification results can be found in **Table S7.**

To generate **Fig. 5**, the participation of most cofactors at a given stratum obtains from its occurrence in a given KEGG line reaction. Cofactors that are not listed in KEGG line reactions (for example PLP-dependent transaminations) were retrieved from SwissProt (release 2025_03) via E.C. numbers associated with the reactions filtered for prokaryotic entries, with subsequent manual curation (**Table S7**). If a target compound was synthesized in more than one reaction in the same stratum, the synthesis was scored with only one bold line in the figure. When both NADH and NADPH could be used in the same reaction, only one occurrence was counted. Reactions that involve the formation or breaking of bonds with phosphate were scored based on **Table S3**.

To generate **Fig. 6**, the reaction network used in the stratification analysis (**Table S7**) was converted into a GML file. Original, non-recoded reactions were used, and $H_2$, $CO_2$, $P_i$, $H_2O$, $NH_3$ and $H_2S$ were set as starting compounds for the figure. Network edges were defined from each reactant (diamond nodes) to the corresponding reactions (circular nodes), and from these reactions to the resulting product compounds (diamond nodes). When the product compound was a chemical moiety carried by cofactors such as tetrahydrofolate, tetrahydromethanopterin, methanofuran or CoA, the cofactor-free, environmental variant was marked in the figure as a node in the corresponding stratum. The resulting network was visualized in Cytoscape (version 3.10.3)[134] and arranged in a bottom-to-top orientation according to the stratification defined in **Table S7**. Annotation data from **Table S4** were integrated, with reactions highlighted in distinct color shades.

*Laboratory reactions and product identification*

Reactions were performed as described in Schlikker *et al.*[30] with the following modifications. Reactors were pressurized with 5 bar argon (99.996%; Messer, Lenzburg, Switzerland). Reaction mixtures were not prepared in a glove box.

**Organic acid and amino acid reactions**. Pyruvate, fumarate, 4-methyl-2-oxopentanoate, and glutamate (20 mM; Merck, Sigma-Aldrich, Darmstadt, Germany) were used as reactants with 0.6 mmol catalyst in a total reaction volume of 1.5 mL. Ni–$SiO_2/Al_2O_3$ (Merck, Sigma-Aldrich, Darmstadt, Germany) was used as catalyst in all reactions; $SiO_2/Al_2O_3$ (Merck, Sigma-Aldrich, Darmstadt, Germany) was additionally included in glutamate cyclization experiments. Reactions were performed at 100 °C and adjusted to pH 9 or 11. Reaction times were 1, 2, 4, or 18 h. Products were identified by $^1H$ NMR spectroscopy at the Center for Molecular and Structural Analytics (CeMSA@HHU, Heinrich Heine University Düsseldorf). $^1H$-NMR spectra were recorded on a Bruker Avance III (600 MHz) spectrometer in a $H_2O/D_2O$ mixture (6/1). Sodium 3-(trimethylsilyl)-1-propanesulfonate (DSS) was used as the internal standard ($CH_3$ peak at 0 p.p.m.) with a noesygppr1d pulse program. The relaxation delay (D1) was 1 s, with a time-domain size (TD) of 98,520 and a sweep width (SWH) of 12,315.271 Hz. We acquired 16 scans per sample. Integration was performed using Chenomx NMR suite (version 9.02). Results are shown in **Fig. S6** (for raw data see **Data availability**).

**Phosphite activation experiments.** Reactions were carried out with 200 mM sodium phosphite ($Na_2HPO_3$; Merck, Sigma-Aldrich) as the reductant. Where indicated, 200 mM ammonium chloride ($NH_4Cl$; Merck, Sigma-Aldrich) was added. Catalysts included nano-nickel ($Ni^0$), nano-iron ($Fe^0$), cobalt powder ($Co^0$), magnetite ($Fe_3O_4$) (all Merck, Sigma-Aldrich), a 1:1 (mol/mol) mixture of iron and nickel (Alfa Aesar) (1.5 mmol per vial), or Pd on activated charcoal (Pd/C) (Merck, Sigma-Aldrich, 0.1 mmol per vial). Reactions were performed at pH 9.0 for 18 h with a total volume of 1.5 mL (1.0 mL for Pd/C reactions). Products were analyzed by $^{31}P$ NMR spectroscopy at the Center for Molecular and Structural Analytics (CeMSA@HHU, Heinrich Heine University Düsseldorf) using a Bruker Avance III 600 MHz spectrometer. Spectra were recorded in a $H_2O/D_2O$ mixture (6/1) with a zgpg30 pulse program. The relaxation delay (D1) was 2 s, with a time-domain size (TD) of 65,536 and a sweep width (SWH) of 96,153.844 Hz. We acquired 16 scans per sample. Spectra were processed with MestReNova (version 14.02) and are shown in **Fig. 4** and **Fig. S4**.

**AMP and serine reactions.** Reactions with adenosine monophosphate (AMP; Merck, Sigma-Aldrich) contained 100 mM AMP and 200 mM sodium phosphite (Merck, Sigma-Aldrich, Darmstadt, Germany), or 200 mM sodium phosphate (Merck, Sigma-Aldrich, Darmstadt, Germany) as control. Selected experiments included 75 mM $NH_4Cl$. Reactions were performed at pH 7.0 (nano-$Ni^0$) or pH 9.0 (Pd/C) and 50 °C with either 1.5 mmol nano-$Ni^0$ or 0.1 mmol Pd/C as catalyst in a 1 mL reaction volume. Reaction times were 4, 18, or 72 h (96 h for $Ni^0$ reactions). ADP formation in Pd/C reactions was analyzed by $^{31}P$ NMR spectroscopy as described above, while ADP over nano-$Ni^0$ was identified by LC-UV-MS (see below).

Reactions with 100 mM serine (Merck, Sigma-Aldrich) were performed with 0.1 mmol Pd/C in a 1 mL reaction volume and 200 mM sodium phosphite under the same conditions. Products were analyzed by $^1H$ NMR spectroscopy as described above. Data are shown in **Fig. 4**, **Fig. S4** and **Fig. S5**.

**ADP identification by LC-UV-MS.** ADP with $Ni^0$ was identified using Liquid chromatography-UV-mass spectrometry with a Dionex UltiMate 3000 UPLC system (Thermo Scientific, Dreieich, Germany) coupled to a maXis 4G (Bruker Daltonics, Bremen, Germany) quadrupole-time-of-flight (Q-TOF) mass spectrometer connected to an electrospray (ESI) ion source. Sample volumes of 10 mL were applied to a 3 mm by 150 mm C18 XSelect® HSS T3 column (2.5 µm particle size, 100 Å pore diameter; Waters, Milford, Massachusetts, USA) and separated using a binary gradient with a flow rate of 0.3 $mL \cdot min^{-1}$. Mobile phase A was water + 0.1% formic acid, and mobile phase B was methanol + 0.1% formic acid. Starting with 5% B, a linear gradient to 95% B was applied from 2.5 to 15 min, followed by 95% B for additional 2 min and return to 5% B within 1 min. The system was equilibrated with 5% B for another 4 min prior to the next injection. The UV detector was set to 254 nm, the wavelength at which adenosine absorbs. The MS (positive-ion mode) was run at 3.5 kV capillary voltage, 1 bar nebulizer pressure, 8 L·min$^{-1}$ dry gas flow, and dry temperature of 200 °C. Data acquisition was performed with COMPASS HYSTAR software (version 6.0.30.0) (Bruker). ADP was identified and quantified from full-scan MS data (mass range 50–1000 m/z) using the DATA ANALYSIS (version 5.3) software (Bruker). Results are shown in **Fig. S5.**

*Statistical analysis*

The significance of the observed archaeal-bacterial distribution pattern was assessed with a two-sided permutation test from the Scipy Python package based on the difference in means of the proportion of genomes carrying the genes (column means in **Table S2**) between bacterial and archaeal taxa, with 1000000 resamples. A p-value < 0.05 was taken as significant.

**Acknowledgements**

We thank the CeMSA@HHU (**Ce**nter for **M**olecular and **S**tructural **A**nalytics @ **H**einrich **H**eine **U**niversity) for recording the NMR-spectroscopic data. Computational infrastructure and support were kindly provided by the Centre for Information and Media Technology at Heinrich Heine University Düsseldorf. JM thanks the NSERC and the Canada Research Chairs Program. We thank Dr. Verena Zimorski and Nils Kapust for help in preparing the manuscript.

**Funding**

ERC grant 101018894 (WFM)

Volkswagen Foundation grant 96_742 (WFM, HT, JM)

Deutsche Forschungsgemeinschaft grant MA 1426/21-1 (WFM)

Deutsche Forschungsgemeinschaft grant TU 315/8–3 (HT)

Spanish Ministry of Science, Innovation and Universities ATRAE grant (HT)

ERC grant 101001752 (JM)

NZ Marsden Fund grant 23-UOC-003 (MS)

**Author contributions**

WFM, NM, NKH, MLS, MBu, LS, CGG, MBr, MS, DHH, MBa, SM, QD, BS, JM, HT and MP conceived and designed the study. WFM, NM, NKH, DHH, MS and LS designed the bioinformatics analysis. NM, NKH, DHH, MS and LS performed the bioinformatics analysis. MLS, MBu, CGG, SM and QD carried out laboratory reactions and analytics. WFM, MLS, MBu, CGG, MBr, SM, QD, BS, JM, HT and MP designed laboratory experiments and interpreted the data. All authors contributed to results discussion. WFM drafted the manuscript. All authors edited the manuscript and approved the final version.

**Competing interests**

The authors declare that they have no competing interests.

**Data availability**

All data needed to evaluate the conclusions in the paper are present in the paper, the Supplementary Materials and in the online repository at https://uni-duesseldorf.sciebo.de/s/jMbfL94nN5FYwBc (for protein families, genome information and experimental raw data). An upgraded version of the CatReNet software which includes the stratification algorithm is available at https://software-ab.cs.uni-tuebingen.de/download/catrenet/welcome.html.

# Supplementary Text

## Reaction stratification: mathematical details

*Definitions and initial results*

A *reaction network* is a set $X$ of molecule types, and a set $R$ of reactions, each represented by writing $r: A \rightarrow B$ where $A$ and $B$ are nonempty subsets of $X$, referred to as the *reactants* and *products* products of $r$ (denoted $\rho(r)$ and $\pi(r)$, respectively). For a set $R'$ of reactions, let $\pi(R')$ be the molecule types that are a product of at least one reaction in $R'$.

We also consider a particular subset of $X$ called the *food set* $F$. A reaction system with a food set is denoted by the triple $(X, R, F)$.

A subset $R'$ of the reaction set $R$ has an *admissible ordering* (or more briefly is *admissible*) if the reactions can be ordered in such a way that each reactant of each reaction $r \in R'$ is either present in the food set or is a product of an earlier reaction in the ordering. In the RAF (Reflexively Autocatalytic and Food set generated) literature (see e.g.[52,133]), this notion is equivalent to the condition that $R'$ is "$F$–generated".

Recall also the notion of the *closure* of the food set under a (sub)set $R'$ of reactions, denoted $cl_{R'}(F)$. This is the set of molecule types correspond to the terminal set $X_n$ of the sequence

$$F = X_0 \subset X_1, \subset \cdots \subset X_n$$

where $X_{i+1}$ is obtained from $X_i$ by adding each molecule type that is not already in $X_i$ and which can be generated by at least one reaction in $R'$ that has all its reactants in $X_i$. The set $cl_{R'}(F)$ can be calculated quickly (i.e. in polynomial time in the size of the system[52].

*The graph $\mathcal{G}(R')$*

Next, consider the directed graph $\mathcal{G}(R')$ that has vertex set $R'$ and with $(r, r')$ being an arc provided that both of the following condition holds:

- At least one reactant $x$ of $r'$ is a product of $r$ and $x \notin cl_{R' \setminus \{r\}}(F)$.

The following lemma is from Steel *et al.*[133] (Lemma 3.1 and Theorem 1).

**Lemma 1**. *In any reaction system with a food set* $(X, R, F)$ *the following four conditions (for any subset* $R'$ *of* R*) are equivalent.*

- *$R'$ has an admissible ordering (i.e. $R'$ is $F$-generated);*

- *$\rho(r) \subseteq cl_{R'}(F)$ for all $r \in R'$;*

- *$cl_{R'}(F) = F \cup \pi(R')$, where $\pi(R') = \cup_{r \in R'} \pi(r)$ (i.e. the set of molecule types generated by the reactions in $R'$).*

- *for each $r \in R', \rho(r) \subseteq \pi(R') \cup F$ and the graph $\mathcal{G}(R')$ has no directed cycles (including any loop on a vertex).*

Next, consider a reaction system with food set, $(X, R, F)$, and a $F$-generated subset $R'$ of $R$. It is possible that a given reaction in $R'$ must always occur before (one or several) other given reactions in $R'$ over all admissible ordering of the reaction set. We describe a mathematical result that allows this ordering information to be determined by a polynomial-time algorithm, and represented graphically.

*An ordering on $R'$*

We define an order on $R'$ by writing $r \prec r'$ if in every admissible ordering of $R'$ reaction $r$ occurs before $r'$.

**Example 1.1:** Consider the following reaction system (which is $F$-generated):

$$r_1: f_1 + f_2 \rightarrow x;$$

$$r_2: f_2 + f_3 \rightarrow y;$$

$$r_3: f_4 + x \rightarrow z,$$

where $X = \{f_1, f_2, f_3, f_4, x, y, z\}$ and $F = \{f_1, f_2, f_3, f_4\}$. Then, $r_1 \prec r_3$.

To determine whether or not $r \prec r'$, let $\xi$ be a new element that is not present in $X$. Given a reaction $r$, let $_{+\xi}r$ be the reaction obtained from $r$ by adding $\xi$ as an additional reactant of $r$. Similarly, given a reaction $r'$, let $r'_{+\xi}$ be the reaction obtained from $r'$ by adding $\xi$ as an additional product of $r'$. Let $R'[_{+\xi}r, r'_{+\xi}]$ be the resulting set of reactions obtained from $R'$ by replacing $r$ with $_{+\xi}r$ and $r'$ with $r'_{+\xi}$, respectively.

**Proposition 1**. *Suppose that $R'$ is an $F$-generated set, and consider any two distinct reactions $r, r'$ in $R'$. The following are then equivalent:*

- *$r \prec r'$.*

- *$R'[{}_{+\xi}r, r'{}_{+\xi}]$ is not $F$-generated.*
- *There is a directed path in $\mathcal{G}(R')$ from $r$ to $r'$.*

*Proof:*

(i) ⇒ (ii): [We show that the negation of (ii) implies (i)] Suppose that the set $R'[{}_{+\xi}r, r'{}_{+\xi}]$ is $F$-generated and let $o'$ be an admissible ordering of this set of reactions. Then $r'{}_{+\xi}$ comes before ${}_{+\xi}r$ in $o'$ (since $r'{}_{+\xi}$ is the only reaction that produces the reactant $\xi$ required for ${}_{+\xi}r$). Moreover, the ordering obtained from $o'$ by replacing ${}_{+\xi}r$ and $r'{}_{+\xi}$ with $r$ and $r'$ respectively is an admissible ordering for $R'$ in which $r'$ comes before $r$ and so $r \prec r'$ does not hold.

(ii) ⇒ (iii): Suppose that $R'[{}_{+\xi}r, r'{}_{+\xi}]$ is not $F$-generated. Then $R'[{}_{+\xi}r, r'{}_{+\xi}]$ violates one of the two conditions stated in Part (iv) of Lemma *1*. Since the first condition in Part (iv) clearly holds for this set of reactions, the second must fail - that is $\mathcal{G}(R'[{}_{+\xi}r, r'{}_{+\xi}])$ contains a directed cycle. However, $R'$ is $F$-generated, and so (again by Lemma *1*) the graph $\mathcal{G}(R')$ has no directed cycle. Moreover, identifying the vertices of these two graphs (i.e. where $r$ is associated with ${}_{+\xi}r$, and $r'$ with $r'{}_{+\xi}$) we see that the only difference between the arc sets of these graphs is that $\mathcal{G}(R'[{}_{+\xi}r, r'{}_{+\xi}])$ contains an arc from $r'{}_{+\xi}$ to ${}_{+\xi}r$, while $\mathcal{G}(R')$ need not have an arc from $r'$ to $r$. Since the former graph has a directed cycle but the latter does not, this means that there is a directed path in $\mathcal{G}(R')$ from $r$ to $r'$, as claimed.

(iii) ⇒ (i): Suppose that $(r_i, r_j)$ is an arc in $\mathcal{G}(R')$. Then in any admissible ordering for $\mathcal{G}(R')$ reaction $r_i$ comes before $r_j$. There is thus a reactant $x$ of $r_j$ that is a product of $r_i$ and, in addition, $x$ is not in the set $cl_{R' \setminus \{r_i\}}(F)$. Thus, in any admissible ordering for $R'$, $r_i$ must come before $r_j$. Therefore, by transitivity, if there is a directed path from $r$ to $r'$ in $\mathcal{G}(R')$, then $r$ comes before $r'$ in any admissible ordering of $R'$, and therefore $r \prec r'$.

**Remark:** The ordering $\prec$ can be extended to include any bidirectional reaction(s) $r$, which has an associated forward reaction $r^+$ and a backward reaction $r^-$. In that case, we write $r_1 \prec r_2$ if $r_1^\alpha \prec r_2^\beta$ for all $\alpha, \beta \in \{+, -\}$.

*Stratifying the reactions and elements in an $F$-generated set*

Given a reaction system with food set $(X, R, F)$ where $R$ is $F$–generated there is a natural way to partition $R$ into 'levels'. Let $R_0 = \{r \in R: \rho(r) \subseteq F\}$, and for $i \geq 1$, let:

$$R_i = \left\{r \in R: \rho(r) \subseteq F \cup \pi\left(\bigcup_{0 \leq j < i} R_j\right)\right\}.$$

By definition, these sets are 'nested' (i.e. $R_i \subseteq R_{i+1}$ for all $i$). Since $R$ is $F$–generated, it follows that $R_i = R$ for some value of $i$ (to see this just consider an admissible ordering of $R$).

For each $r \in R$, we can then define an associated ranking $\lambda(r)$ to be the smallest value of $j$ for which $r \in R_j$. Thus the collection of reactions in $R$ with values $\lambda(r) = 0,1,2 \ldots$ partitions $R$ into disjoint subsets. Suppose that $R$ is an $F$-generated set for $(X, R, F)$, and $r, r' \in R$. Then:

$$r \prec r' \Rightarrow \lambda(r) < \lambda(r').$$

To justify this claim, we establish the contrapositive. Suppose that $\lambda(r') \leq \lambda(r)$. Then there is an admissible ordering of $R$ in which $r'$ comes before $r$. Thus, not every admissible ordering of $R'$ has $r$ before $r'$, so $r \prec r'$ does not hold.

Note also that since $R$ is $F$–generated and we write $r < r'$ if $r \neq r$ and at least one reactant of $r'$ is a product of $r$ but is not a product of any other reaction in $R$ (e.g., in Example 1.1 we have $r_1 < r_3$), then the condition $r < r'$ implies that $r \prec r'$.

In a similar manner, we can also obtain a stratification of $\Pi(R)$ for any $F$–generated set $R$, as follows. Let $X_0 = F$, and for each $i \geq 1$, let:

$$X_i = X_{i-1} \cup \left( \bigcup_{r \in R: \rho(r) \subseteq X_{i-1}} \pi(r) \right).$$

Thus $X_i$ is $X_{i-1}$ together with all the products of all the reactions in $R$ that have all their reactants in $X_{i-1}$. The resulting collection of sets of elements from $X$ provides a nested increasing sequence of sets $X_0 \subset \cdots \subset X_K = X_{K+1}$, where $X_K = cl_R(F)$. Next, define $\lambda: \pi(R) \to \{1,2, \ldots\}$ as follows:

$$\lambda(x) = \min\{i: x \in X_i\}.$$

In this way, the elements of $\pi(R)$ are partitioned by their $\lambda$ values. Moreover, the function $\lambda$ has the following interpretation: $\lambda(x)$ is the length of the shortest admissible ordering of reactions (from $R$) that ends in a reaction that has $x$ as a product. Thus if $\lambda(x) < \lambda(y)$, the element $x$ can arise in fewer admissible 'steps' from the food set $F$ than element $y$ can.

The stratification of reactions and molecule types has been implemented into the open-source software package *CatReNet*[132] for use in this paper.

**Structures and taxonomic distribution of complexes outside the biosynthetic core**

*Gene distributions*

Protein families obtained as described in **Methods** were not annotated with sequences for energy conservation enzymes, as they are not part of the autotrophic biosynthetic core. In order to obtain the taxonomic distributions of the Mtr and the Rnf complex (**Fig. S3, Table S6**), one query sequence for each subunit of the complexes was retrieved (GenBank: AJ132817.2[135] and GenBank: FJ416148.1[136], respectively). For ATP synthase subunits, all SwissProt prokaryotic sequences that could be identified by a combination of the gene name (**Table S6**) and "ATP synthase" or "ATPase" in the protein name, were retrieved. These sequences were used as queries in DIAMOND blastp alignments against the balanced dataset of 401 archaeal and 552 bacterial genomes (see **Methods**). BLAST hits with a local sequence identity of ≥25% and an e-value of $\leq 10^{-10}$ were considered to be homologs.

The taxonomic distribution of homologs for each subunit of Mtr and Rnf is shown in **Table S6 and Fig. S3**. The taxonomic distribution of subunits for bacterial (F-type, F-ATPase) and archaeal-type (V-type or V-ATPase) ATP synthases is shown in **Fig. S3**. Universal and domain-specific ATPase subunits were identified based on the literature[137]. When BLAST searches were not sensitive enough to detect homologies known from the structural literature, rows were manually joined to reflect known evolutionary relationships (**Table S6**).

*Protein structure visualization*

Figures of protein structures were rendered with Pymol (The PyMol Molecular Graphics System, Version 2.5.4., Schrödinger, LLC). For ATP synthases (PDB ID: 6OQR[138]; PDB ID: 6QUM[139]) domain-specific components were colored according to[128–130,137,140]. The Rnf structure (PDB ID: 7ZC6[141]) is missing the small mobile FdI domain. The Mtr structure (PDB ID: 8Q3V[142]) is missing subunits MtrH and $MtrA_s$. They were added for **Fig. S3** similarly as in[142]. The complex $Mtr(A_cBGF)_3H_2$ modelled with AlphaFold 3[40] was superimposed with the experimental structure. MtrH was repositioned to reflect the orientation in[142], and superimposed with the $MtrA_sH_2$ complex modelled with AlphaFold 3 to position $MtrA_s$. The cobalamin cofactor was modelled from a superimposed experimental structure of MtrA with cobalamin (PDB ID: 5LAA[143]. Archaeal transmembrane complexes were embedded in a glycerol-dialkyl-glycerol tetraether (GDGT)[140] lipid monolayer, while bacterial transmembrane complexes were embedded in a palmitoyl-oleoyl-phosphatidylethanolamine (POPE)[145] lipid bilayer. Membrane positioning is schematic.

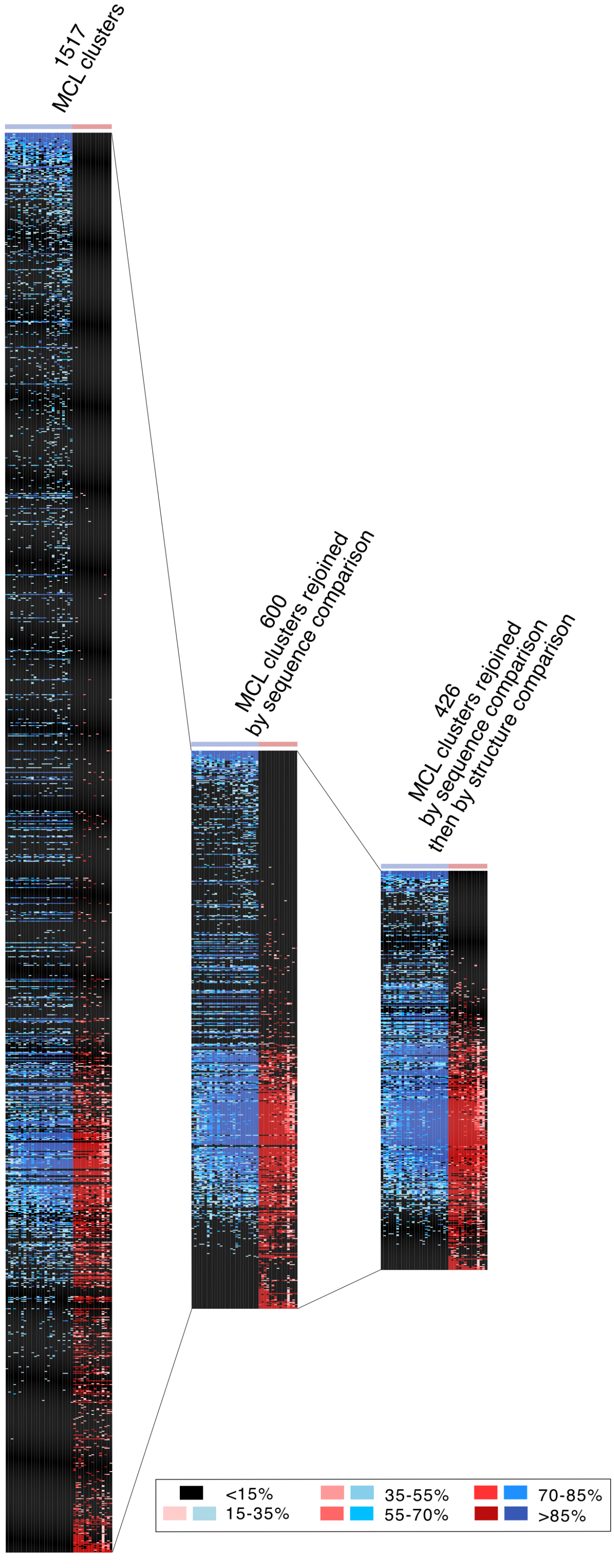

**Fig. S1. Merging of core metabolism protein families based on sequence and structural similarity**. The distribution of protein families across bacterial classes (blue ticks) and archaeal orders (red ticks) is shown for raw sequence-based MCL clusters (left), clusters after sequence similarity joining (center) and clusters after subsequent structure similarity joining (right)

(**Table S9**). Only clusters that mapped to the same KEGG reaction were checked against joining criteria (**Methods**). Color intensity indicates the proportion of genomes per taxon possessing the gene (scale at the bottom). Taxa order is the same as in **Fig. 1**.

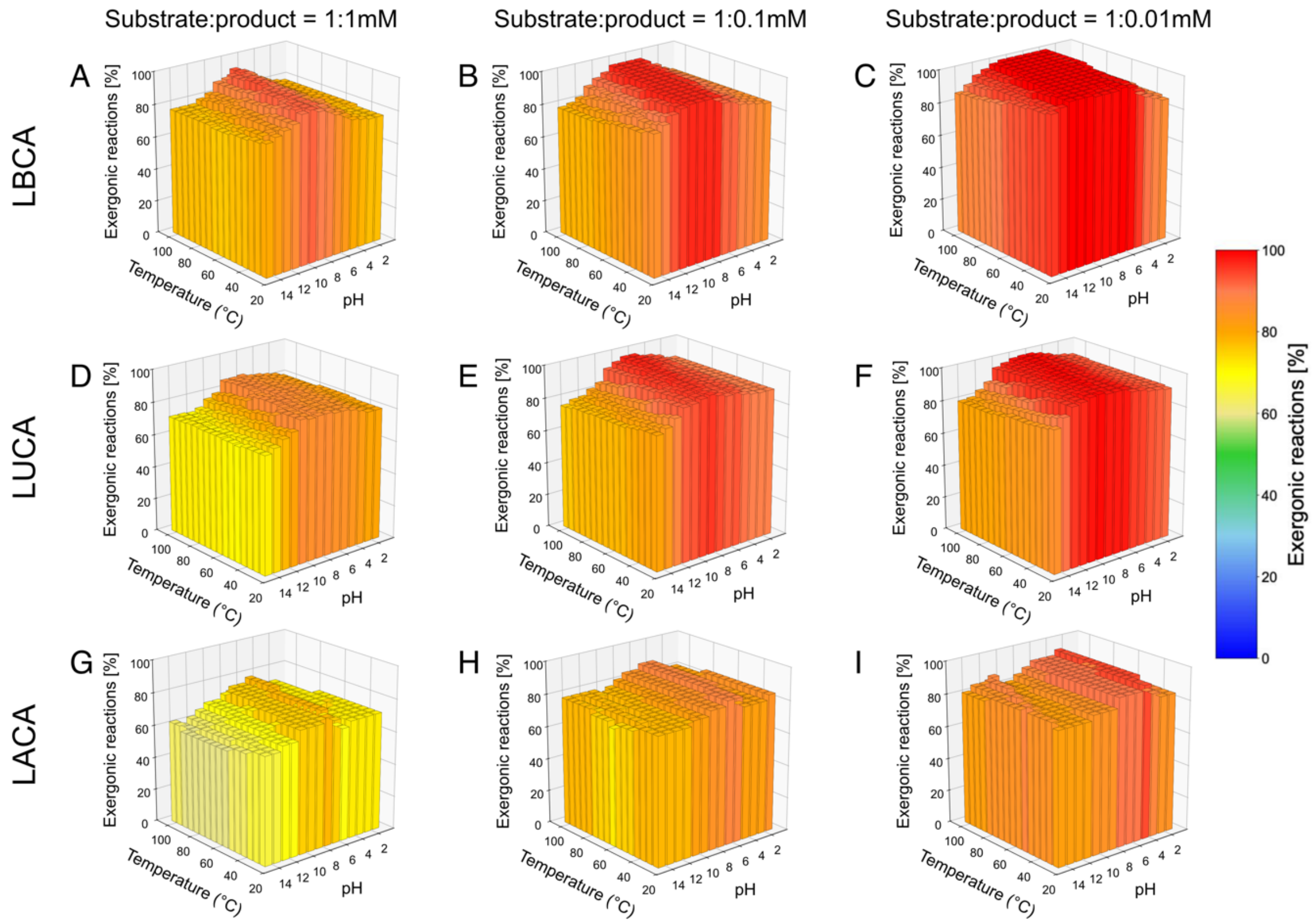

**Fig. S2. Proportions of exergonic reactions in LBCA, LUCA and LACA**. Proportions of exergonic and close-to-equilibrium reactions ($\Delta G$ < 5 kJ/mol) are plotted for each set of conditions for 326 reactions that yielded a $\Delta G$ value. Ionic strength (250 mM) and reactant concentration (1 mM) are constant, while product concentration is decreased from 1mM (**A**, **D**, **G**) to 0.1 mM (**B**, **E**, **H**) and 0.01 mM (**C**, **F**, **I**). Plots for reactions assigned to LBCA are shown in **A**, **B** and **C**, plots for reactions assigned to LUCA are in **D**, **E** and **F**, while reactions assigned to LACA are shown in **G**, **H** and **I**. Grey zones from **Fig. 1B** were excluded. Underlying data can be found in **Table S5**.

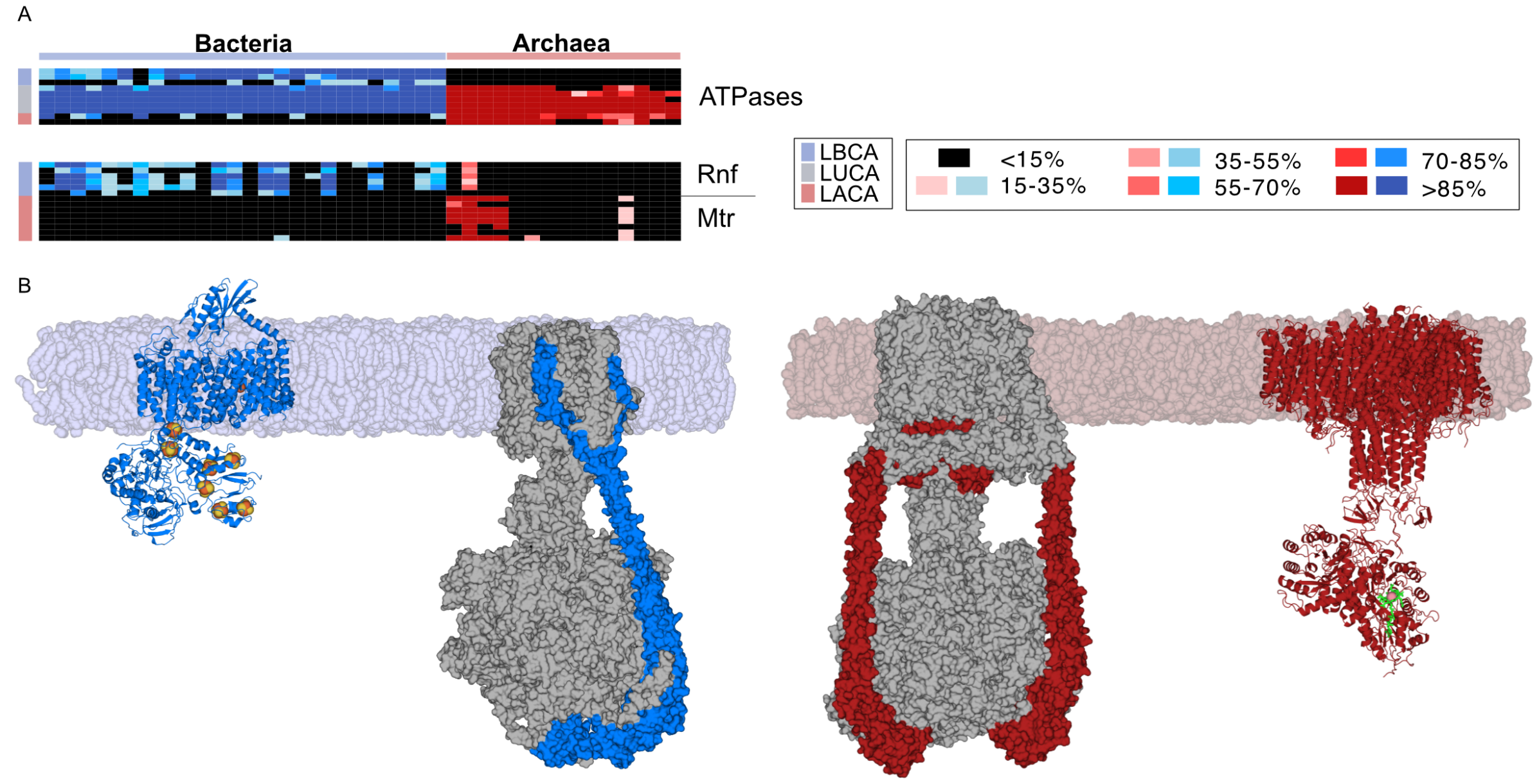

**Fig. S3. Distribution and protein structures of ancestral energy conservation enzymes in bacteria and archaea**. For escape as free-living cells, metabolism had to achieve the state of a balanced stoichiometric reaction[145], called redox balance in physiology, and the exergonic reactions of electron transfer from $H_2$ to $CO_2$ needed to be coupled to the pumping of ions from the inside of the cell to the outside (chemiosmotic coupling), as in modern cells. Prior to the coupling of $CO_2$ reduction to ion pumping, the natural pH gradient provided by serpentinizing systems (pH 9–11, vs. sea water (pH 6.5[146]) was harnessable with the help of simple, abiotically synthesized hydrophobics[147,148] such as fatty acids, which were recently shown to support the ATP-synthesizing function of a rotor-stator ATPase[149]. **A**. The distribution pattern of genes for ATPase subunits resembles ribosomal proteins (**Fig. 1A**) and core metabolism (**Fig. 1B**). The ATPase needed to harness geochemically generated ion gradients is an invention of the protein world and has no geochemical analogue. In order to replace the natural pH gradient by a biological mechanism, much like replacement of solid-state metals by cofactors, enzyme innovations were required. Gradient-generating pumping complexes originated separately in the bacterial (Rnf) and archaeal (Mtr) domains. The occurrence of Rnf in Methanosarcinales, the most recent order of methanogens[150], is known and likely due to lateral gene transfer from bacteria[151]. Genes present in bacterial genomes are plotted as blue ticks, while those in archaeal genomes are in red. Color intensity indicates the proportion of genomes per taxon possessing the gene, according to the scale in **Fig. 1**. For details see **Table S6**. **B**. Structures of the ATP synthase[138,139], Rnf[141] and Mtr[40,142,143] are shown embedded in bacterial and archaeal lipids. Before the independent origin of lipid synthesis in bacteria and archaea, the ATP synthase could have functioned in simple abiotically synthesized fatty acid membranes[149]. ATP synthase

subunits that arose independently in the two domains are shown in red (archaea) and blue (bacteria)[137]. The cobalamin cofactor in Mtr is shown in green. FeS clusters are highlighted.

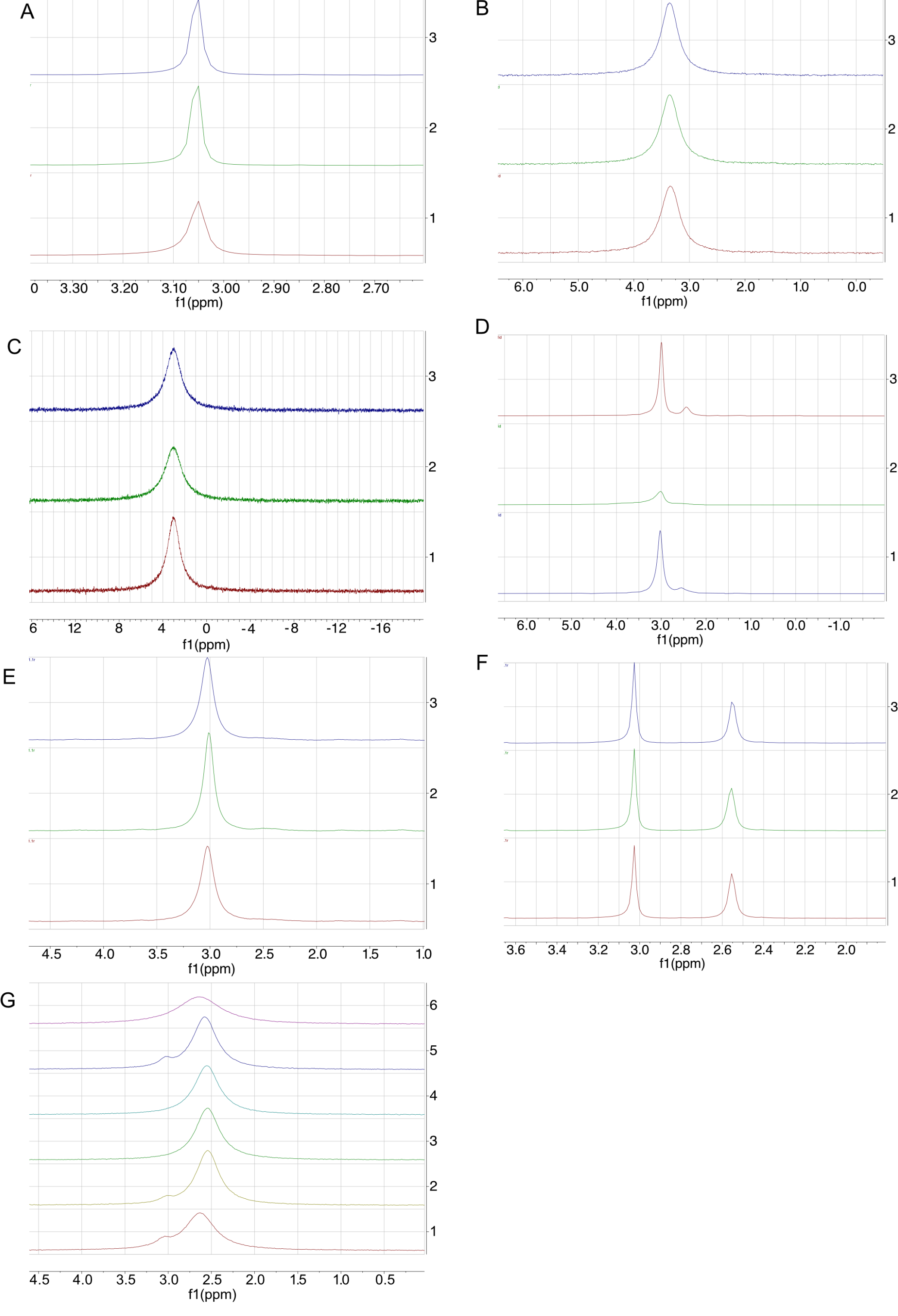

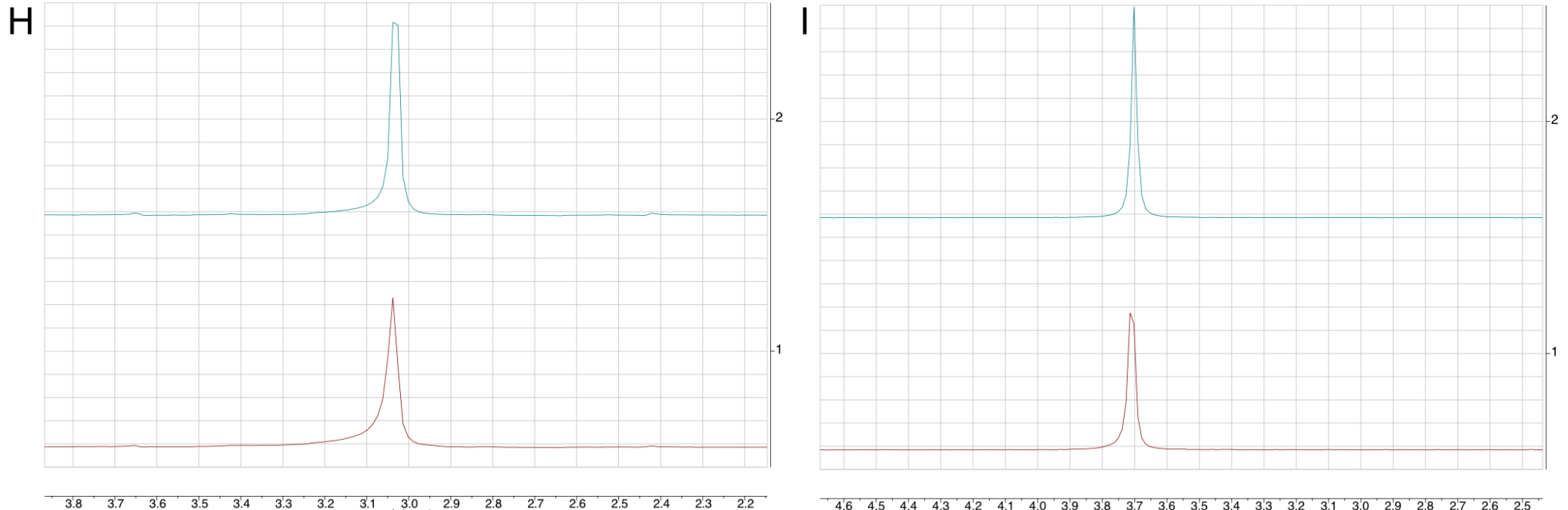

J

| **Phosphite** | **Y = 2654577*X** | | | | | |
|---|---|---|---|---|---|---|
| Number | A 1 | A 2 | A 3 | B 1 | B 2 | B 3 |
| Peak area | 504904842.8 | 537129478.3 | 501256867.1 | 515612407.6 | 520742115.9 | 482383474.6 |
| Range | 3.30 .. 2.79 | 3.45 .. 2.88 | 3.48 .. 2.72 | 6.50 .. 1.08 | 6.84 .. 0.78 | 5.08 .. 2.04 |
| Number | C1 | C 2 | C 3 | D 1 | D 2 | D 3 |
| Peak area | 523528584.1 | 501287103.1 | 548024242.5 | 307134782.5 | 418839145.6 | 294368170.6 |
| Range | 10.34 .. -3.89 | 11.80 .. -5.02 | 7.93 .. -2.07 | 4.32 .. 2.65 | 5.01 .. 2.78 | 4.25 .. 2.72 |
| Number | E 1 | E 2 | E 3 | F 1 | F 2 | F 3 |
| Peak area | 530068031 | 514601543.3 | 524254441.8 | 282730340.3 | 294682964.3 | 251745488 |
| Range | 4.23 .. 2.01 | 4.13 .. 2.08 | 4.13 .. 1.98 | 3.61 .. 2.84 | 3.51 .. 2.80 | 3.31 .. 2.80 |
| Number | G 1 | G 2 | G 3 | G 4 | G 5 | G 6 |
| Peak area | 56483887.06 | 95561989.06 | 50162016.31 | 40832622 | 52254441.75 | 133123559.3 |
| Range | 4.04 .. 2.95 | 4.48 .. 2.90 | 4.83 .. 2.83 | 4.36 .. 3.35 | 4.79 .. 2.95 | 4.95 .. 2.96 |

K

| **Phosphate** | **Y = 1363794*X** | | | | | |
|---|---|---|---|---|---|---|
| Number | D 1 | D 2 | D 3 | F 1 | F 2 | F 3 |
| Peak area | 120172930.9 | 37382853.5 | 96519876.62 | 136877987 | 152772728.3 | 143981088.5 |
| Range | 2.67 .. 1.37 | 2.78 .. 1.84 | 2.72 .. 1.98 | 2.84 .. 2.06 | 2.81 .. 2.26 | 2.79 .. 2.22 |
| Number | G 1 | G 2 | G 3 | G 4 | G 5 | G 6 |
| Peak area | 231528831.1 | 219071578.5 | 257161692.8 | 254339894.3 | 220315705.5 | 211664539.8 |
| Range | 5.41 .. -1.43 | 2.93 .. 0.34 | 4.83 .. 0.80 | 4.58 .. 0.36 | 2.97 .. -0.20 | 2.98 .. 0.39 |
| Number | H 1 | H 2 | I 1 | I 2 | | |
| Peak area | 282592437 | 264533092 | 273910446.8 | 285342174.3 | | |
| Range | 1.40 .. 1.15 | 1.51 .. 1.11 | 1.90 .. 1.53 | 1.86 .. 1.54 | | |

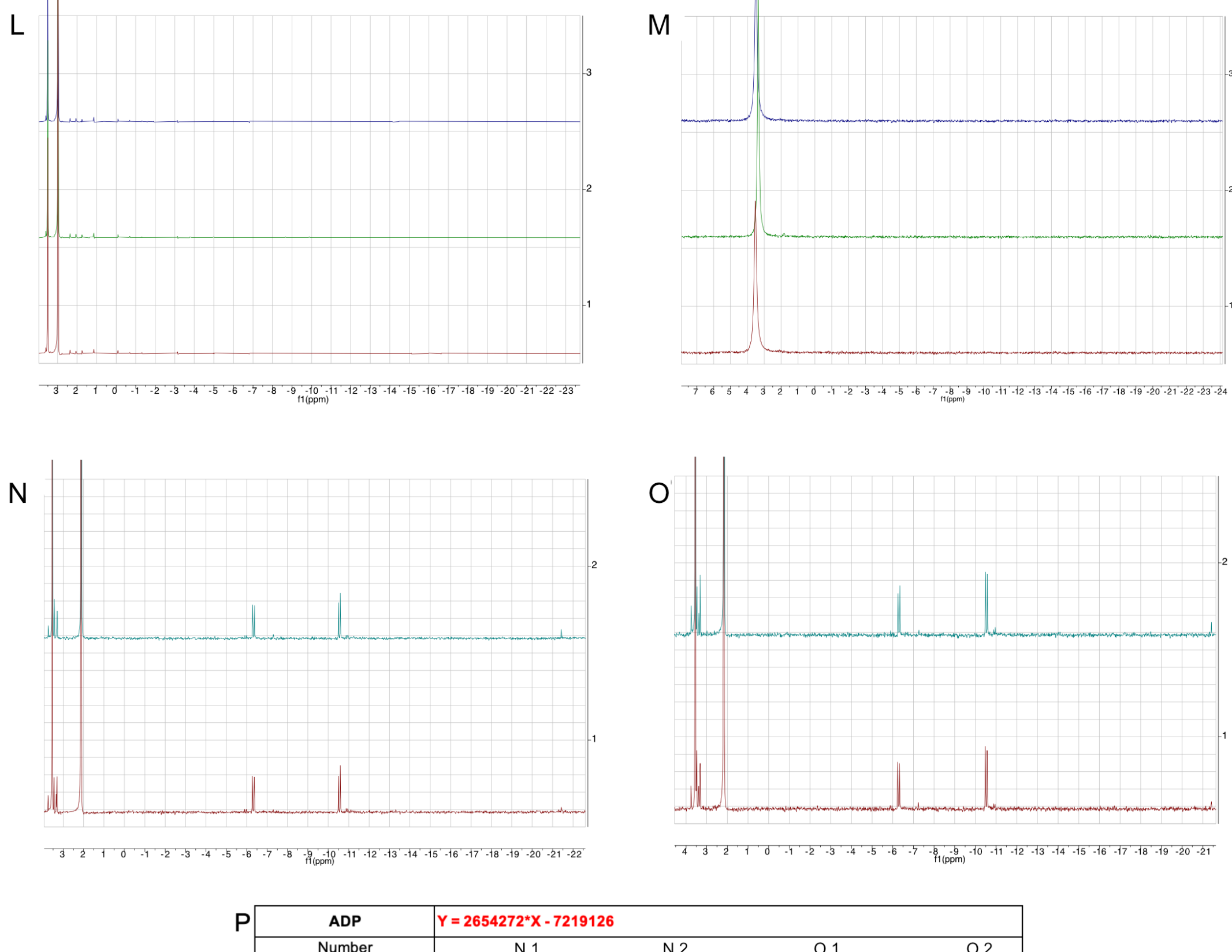

P

| ADP | Y = 2654272*X - 7219126 | | | |
|---|---|---|---|---|
| Number | N 1 | N 2 | O 1 | O 2 |
| Peak area | 13279062.25 | 13020027.75 | 15083206.25 | 14427535.5 |
| Range | -10.38 .. -10.70 | -10.33 .. -10.76 | -10.38 .. -10.75 | -10.40 .. -10.70 |

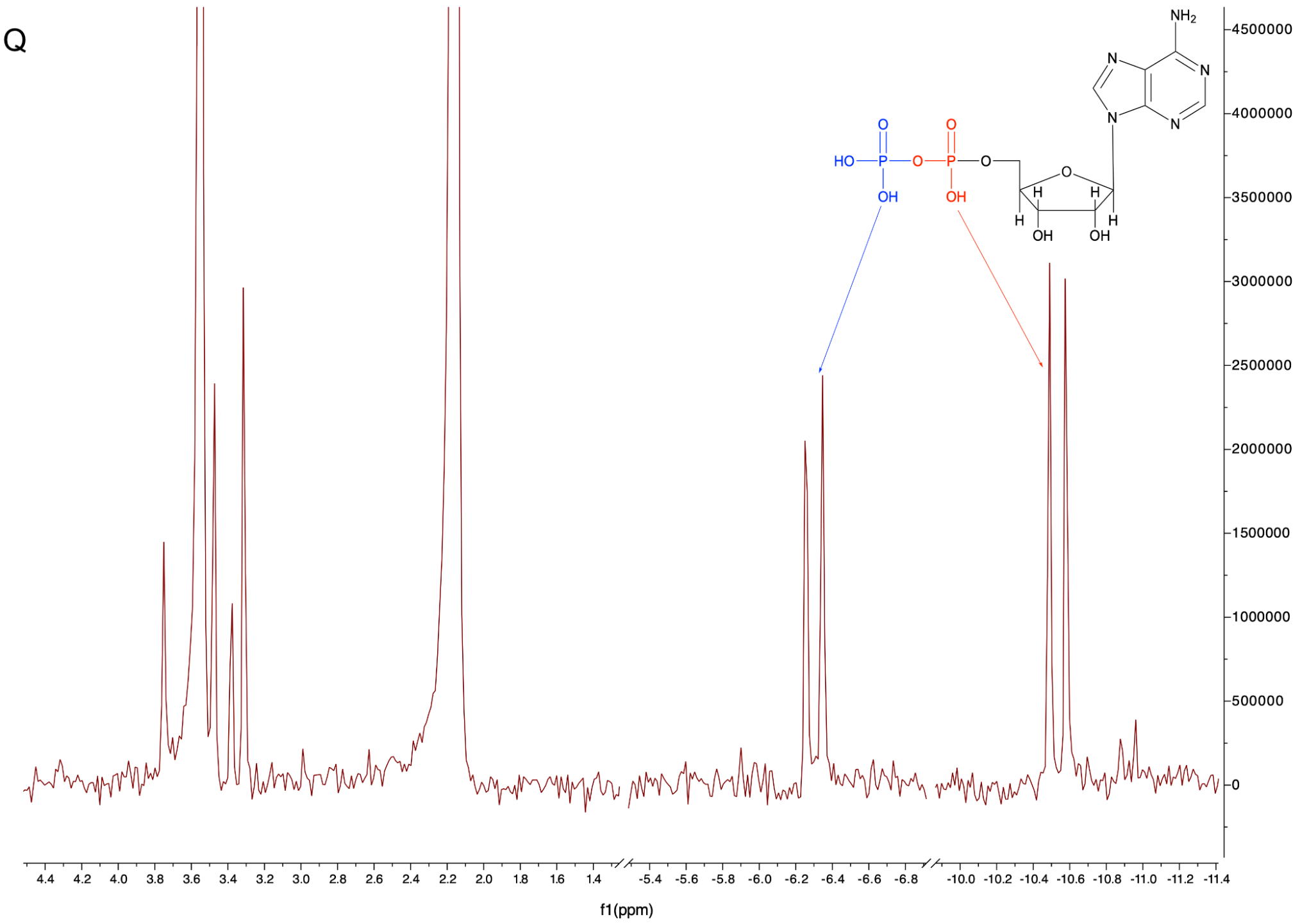

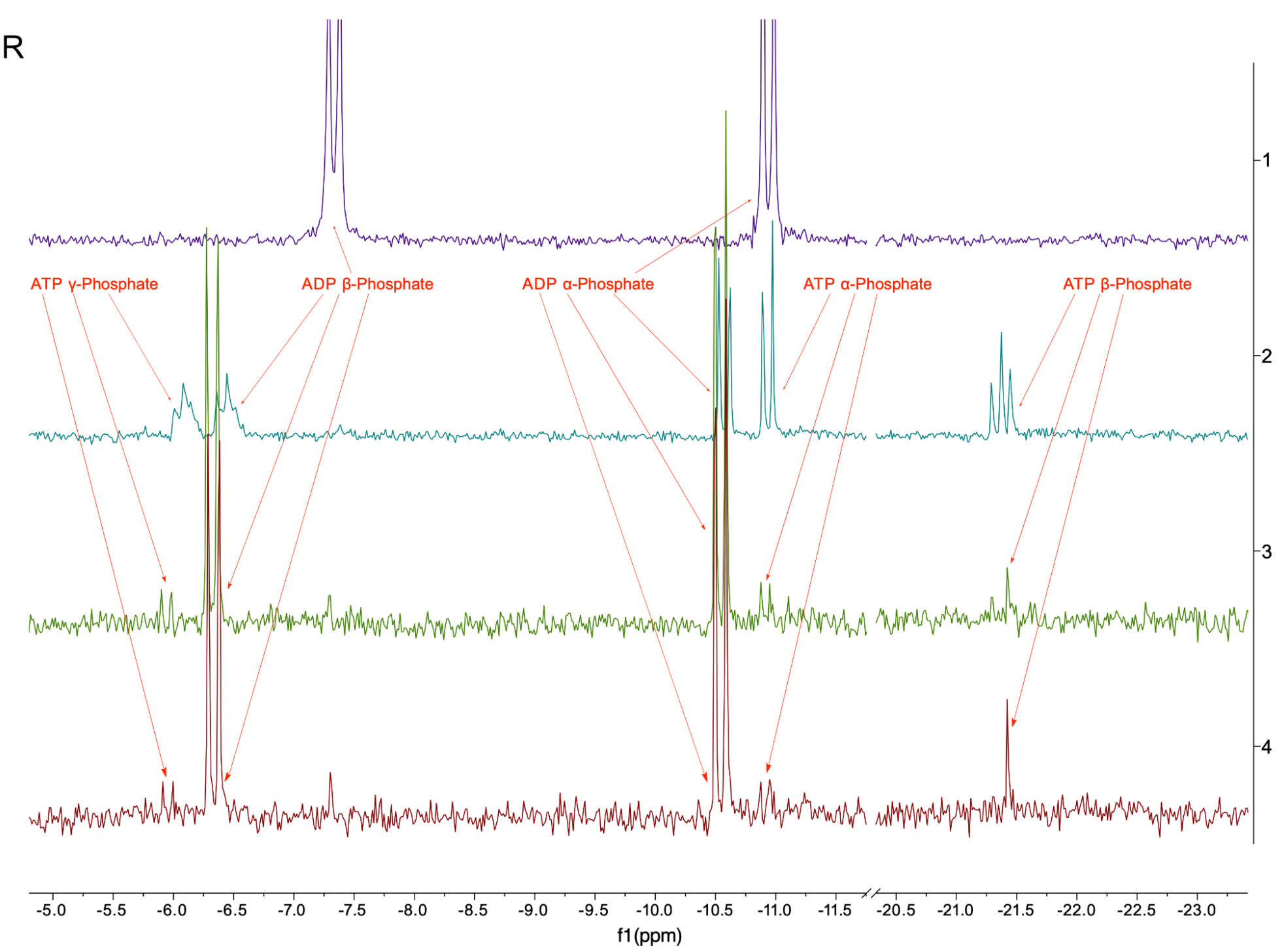

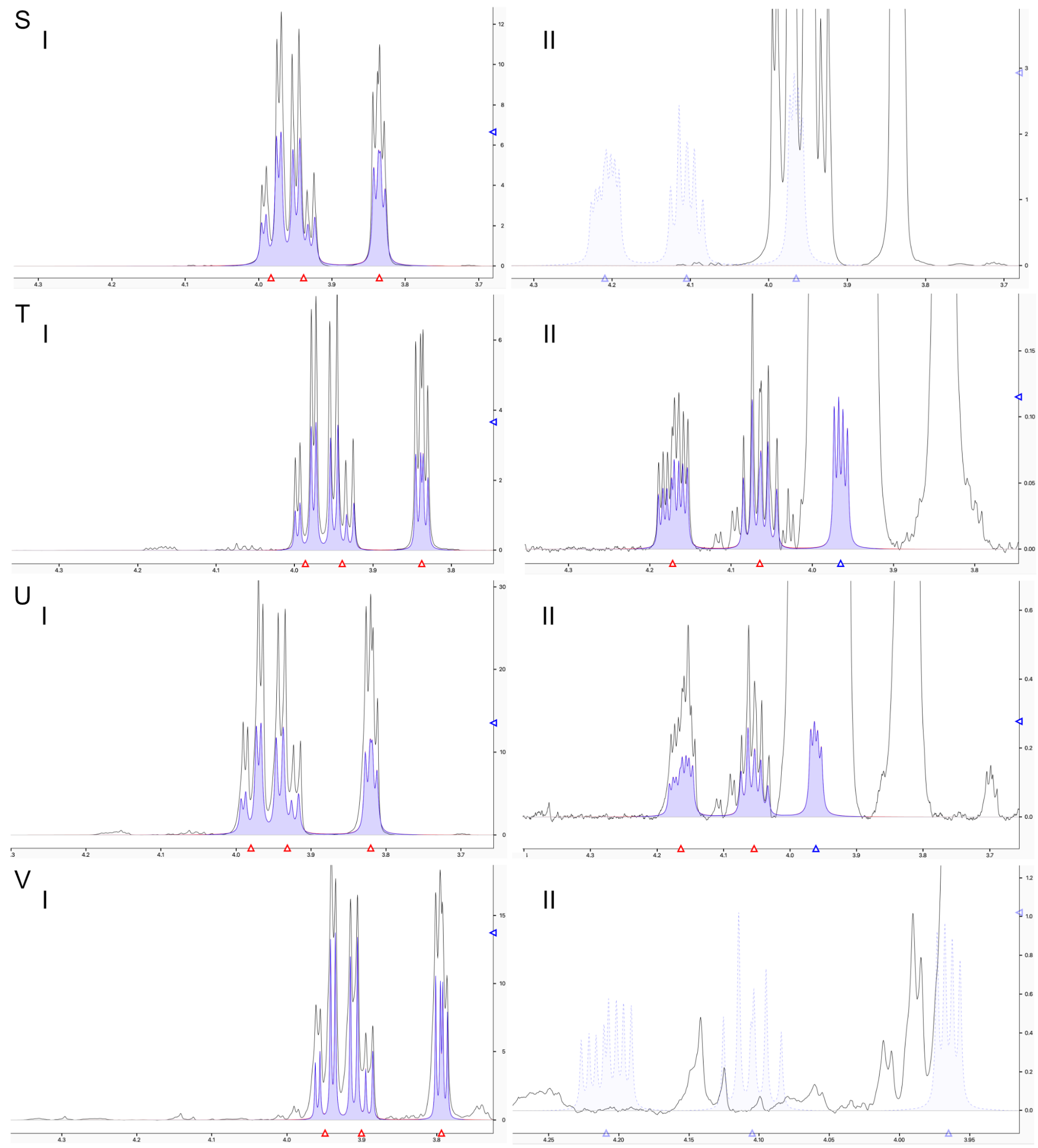

**Fig. S4. A-I** Raw spectral data ($^{31}$P NMR) for **Fig. 4A** in appearing order. **J-K,** Integrals and measured ranges for the specific peaks A-I. Phosphite and phosphate were identified by comparison to a concentration series and the corresponding calibration curve (shown in red). **L-O**, Raw spectral data ($^{31}$P NMR) related to AMP phosphorylation products corresponding to **Fig. 4B** (left) shown in the same order as in the main figure. **P**, Integrals and measured ranges for the specific peaks L-O. The calibration curve from a concentration series is shown in red. **Q**, Phosphorus assignment of ADP signals in the $^{31}$P NMR spectra. **R**, ADP and ATP peak assignments in the $^{31}$P NMR spectra. Spectrum 1 shows the ADP standard. Spectra 3 and 4 display the unmodified sample, showing ADP accumulation and low-level ATP accumulation, whereas Spectrum 2 shows the same sample (Spectrum 3) after spiking with ATP. The peaks specific for α-, β- and γ- phosphates of ATP are indicated. Note that $^{31}$P NMR peaks undergo pH-dependent shifts. **S-V**, Raw spectral data ($^{1}$H NMR) of serine (**I**) and phosphoserine (**II**)

corresponding to **Fig. 4B** (right). The evaluation was performed using the internal Chenomx software library. Spectra are shown in pairs (**S**/**T**, **U**/**V**, **W**/**X**, **Y**/**Z**) for direct comparison of serine and phosphoserine. (**S**, **U**, **W**, **Y**), Blue spectra represent serine standards, while black spectra represent reaction samples. (**T**, **V**, **X**, **Z**), The same reaction samples overlaid with phosphoserine standards. Dashed blue lines indicate the absence of the respective compound. All reaction conditions are described in **Methods**. Nagaosa and Aoyama[152] reported efficient phosphite oxidation using 10% Pd on carbon in the presence of $O_2$, they suggested a mechanism involving $H_2O_2$. Our reactions were performed under Ar. In our phosphite to phosphate conversion reactions, the pressure in the reactor (0.5 L volume) increases slightly by approx. 0.5 bar, depending upon the reaction, corresponding to the accumulation of 0.05 L of gas, corresponding to roughly 2.2 mmol of $H_2$ gas as a reaction product; a typical reactor with 10.5 ml total volume of 200 mM phosphite would generate, under complete phosphite oxidation, roughly 2.1 mmol $H_2$, indicating a phosphite activation reaction (for fully protonated species) according to $H_3P^{+III}O_3 + H_2O \rightarrow H_3P^{+V}O_4 + H_2$.

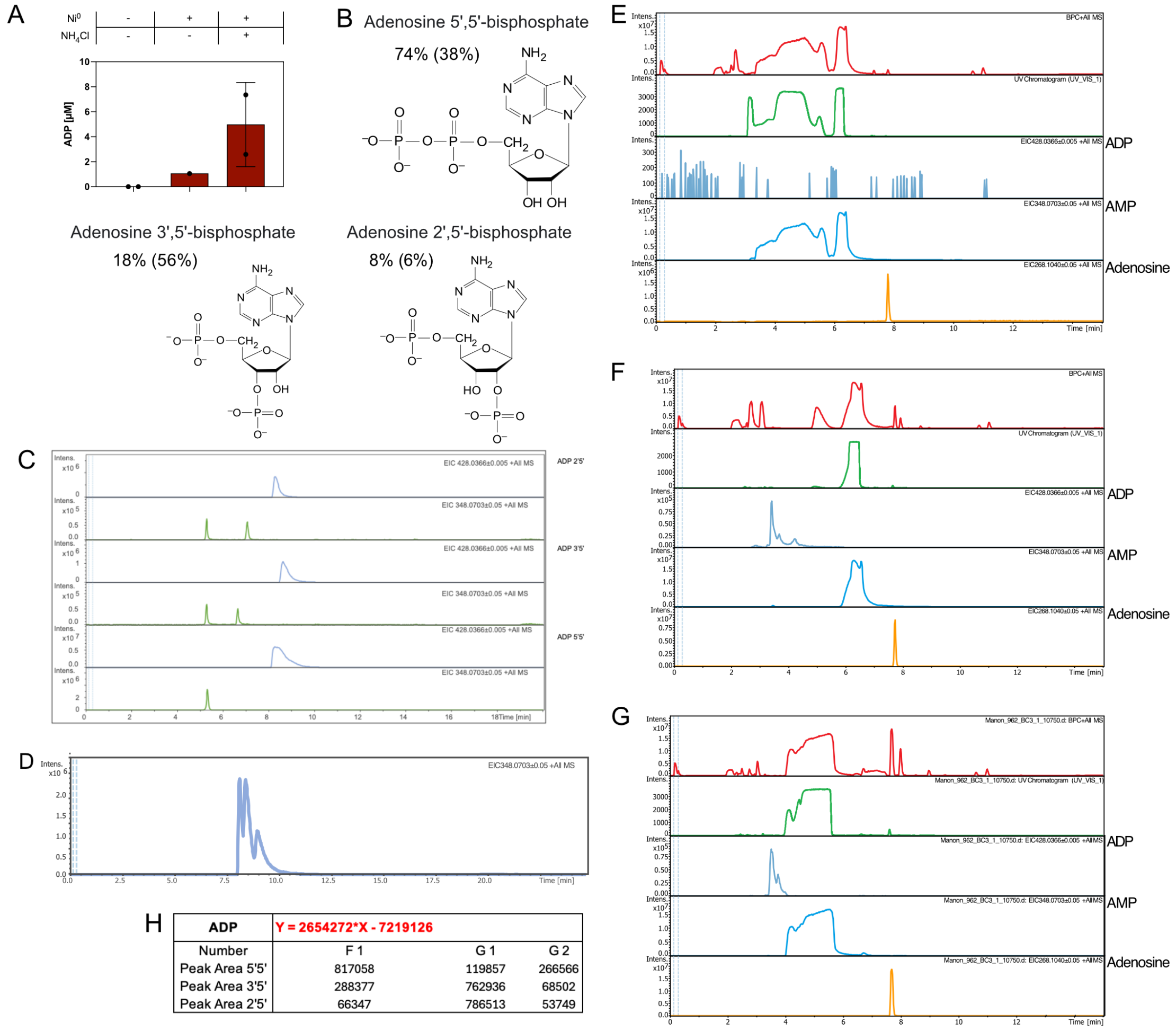

| ADP | Y = 2654272*X - 7219126 | | |
|---|---|---|---|
| Number | F 1 | G 1 | G 2 |
| Peak Area 5'5' | 817058 | 119857 | 266566 |
| Peak Area 3'5' | 288377 | 762936 | 68502 |
| Peak Area 2'5' | 66347 | 786513 | 53749 |

**Fig. S5. A–H AMP phosphorylation to ADP over Ni$^0$. A**, Concentration of ADP formed from AMP in the presence or absence of $NH_4Cl$ and Ni$^0$ in a 96 h reaction at pH 7 and 50 °C. **B**, Structures of different ADP isomers. Percentage values outside parentheses indicate the proportions of ADP isomers in the absence of $NH_4Cl$, while values in parentheses refer to reactions conducted with $NH_4Cl$. **C**, LC–MS chromatograms of ADP isomer standards. **D**, Extracted ion chromatograms (EICs) of the three ADP isomers obtained from AMP phosphorylation reactions. **E-G**, LC–MS analysis of AMP phosphorylation reactions corresponding to the first (**E**), middle (**F**), and last (**G**) conditions shown in panel A. Each panel displays extracted ion chromatograms (EICs) for ADP, AMP, and adenosine, as well as UV and base peak chromatograms. **H**, Calibration curve (red) and corresponding peak areas for the different ADP isomers (5′5′, 3′5′, 2′5′).

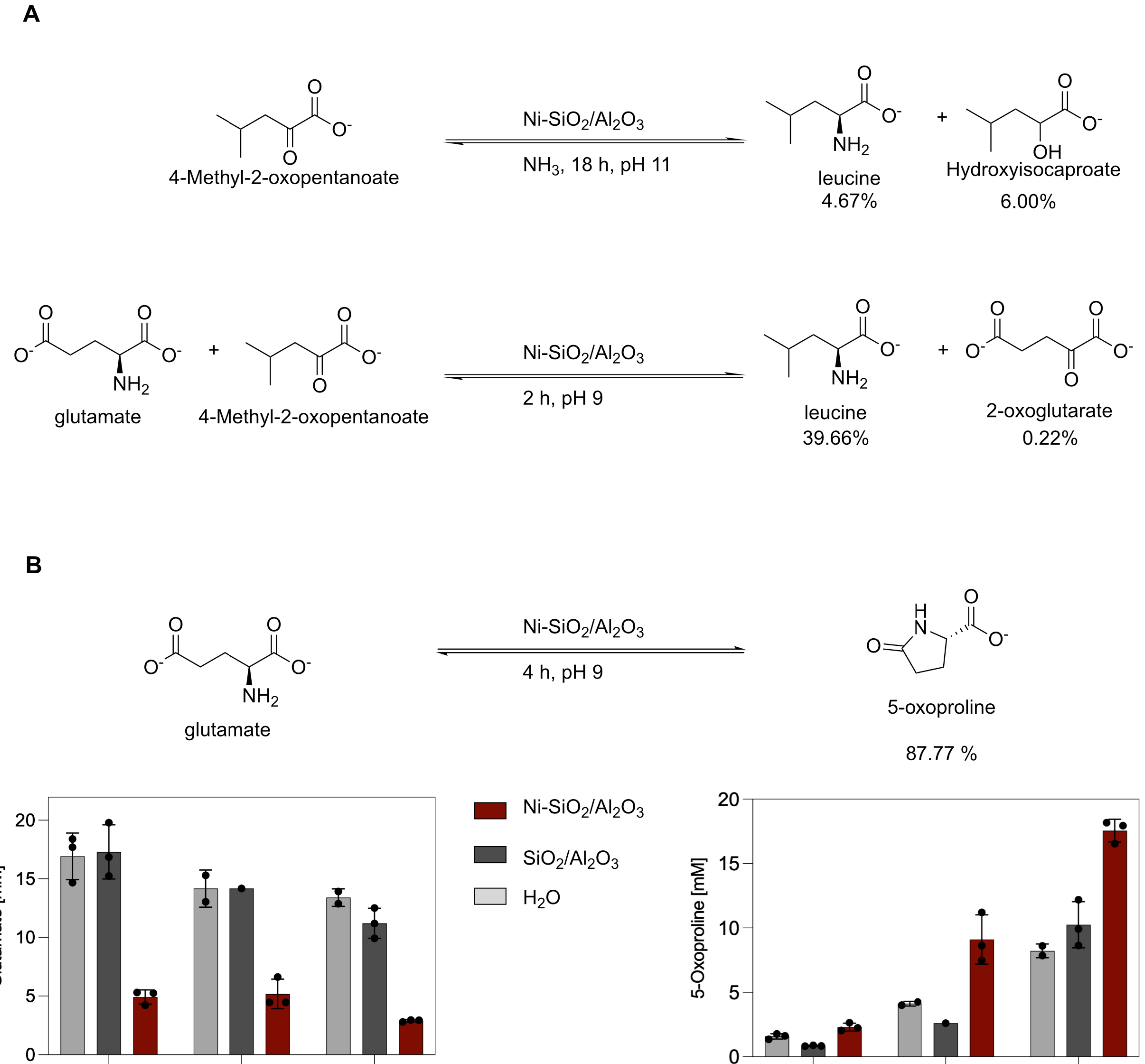

**Fig. S6. Reactions experimentally shown in this work under simulated hydrothermal conditions. A.** Core metabolism reactions with their associated KEGG identifiers. **B.** Intramolecular cyclization of glutamate into 2-oxoproline (pyroglutamate) with amide bond formation. Concentrations of the reactant (glutamate) and product (2-oxoproline) are plotted after a reaction time of 1 h, 2 h, and 4 h with different catalysts. Reaction conditions are described in **Methods**, with specific conditions marked underneath the arrows. Yields are shown underneath each product.

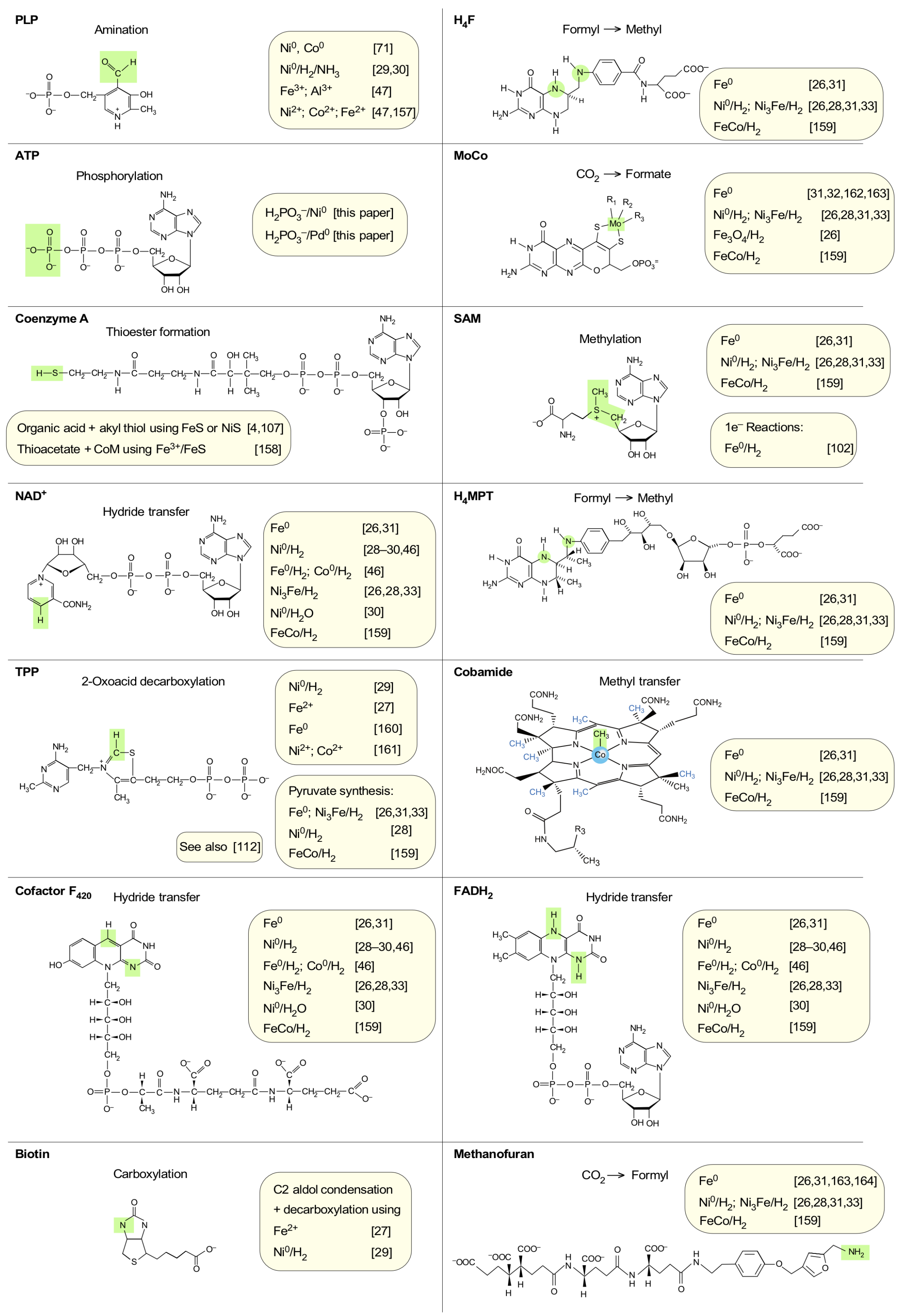

**Fig. S7. Functions of cofactors that are replaced by metals.** Metals that catalyze the reaction are indicated in shaded boxes. References are indicated in brackets. PLP can perform a number of functions performed by TPP[112] and is likely the evolutionary precursor of TPP[112]. Water is the most common reactant in metabolism, indicating that metabolism arose in the aqueous

phase, but studies employing gas phase reactions achieve $CO_2$ fixation using $H_2$ and $Co^0$, $Fe^0$, and alloys[153] or ultraviolet-light enhanced $CO_2$ fixation over metal sulfide catalysts[154]. For the conversions cited in the figure, catalysts are essential. In the absence of catalysts, $CO_2$ conversion to formate or methane is negligible, even after 5 years at 300°C and 350 bar[155], whereas in the presence of native metal catalysis, 0.1–0.5 M formate can be produced in water from $H_2$ and $CO_2$ gas overnight[26,28,31,33] (reviewed[32]). Ion gradients are, however, not required for any of the reactions shown here. It has been reported that an ion gradient can convert $CO_2$ to 1 μM formate[156], yet 200,000 times higher formate concentrations (in addition to C2 and C3 products) are obtained from $H_2$ and $CO_2$ using of $Ni^0$ or $Ni_3Fe$[26,28,31,33].

**Table S1. (separate file) Distribution of ribosomal proteins across prokaryotic higher taxa.**

The file contains the distribution of 83 ribosomal proteins across bacterial classes and archaeal orders and a list of ribosomal proteins from *E. coli* and *T. kodakarensis* used to retrieve ribosomal protein sequences. Universal and domain-specific ribosomal proteins are marked.

**Table S2. (separate file) Distribution of core metabolic enzymes across prokaryotic higher taxa and substrate generality assignment.**

The file contains core biosynthetic enzyme distributions across bacterial classes and archaeal orders and their assignment to LUCA, LBCA or LACA. Additionally, the overlap with KOs identified in a recent genome-wide inference of LUCA based on phylogeny from Moody *et al.*[20] and literature references for the characterization of LACA and LBCA enzymes are listed. A sheet with the assignment of substrate generalists (enzymes with broad substrate specificity) in the biosynthetic core is also provided. Substrate generality is traditionally understood as an essential property of primordial enzymes, as it reduces the number of enzymes that the first cells required. The generality of enzymes (accepting many substrates) and cofactors (each used by many enzymes) was preceded in metabolic assembly by substrate generality of metabolism's metallic inorganic precursors.

**Table S3. (separate file) Types of chemical reactions catalyzed by core metabolism protein families.**

Reaction equations and assignments of reaction types are given for 361 reactions assigned to protein families.

**Table S4. (separate file) Core metabolic reactions reported experimentally under serpentinizing vent conditions.**
The file contains a referenced list of core metabolic reactions mapped to protein families that have been experimentally shown to proceed under hydrothermal conditions, without catalysis or catalyzed by transition metals. Reactions shown to occur on FeS minerals, Fe(II/III) oxyhydroxides or with the use of an external electric current are marked in a separate column. The reactions shown experimentally are assigned to three categories: A, the reaction has been reported experimentally in hydrothermal conditions; B, the same chemical transformation on a different substrate has been reported experimentally; C, the reaction or pathway it belongs to has been reported experimentally, even though the pathway from reactants to products in the laboratory involves a different number of steps or a different reaction sequence compared to metabolism.

**Table S5. (separate file) Gibbs energy ΔG for core metabolic reactions.**
Gibbs energy values (ΔG) of core metabolic reactions are given across a range of pH, temperatures and concentrations. Only values for reactions mapped to protein families that yielded a ΔG value are listed.

**Table S6. (separate file) Distribution of ATPase, Mtr and Rnf subunits across prokaryotic higher taxa.**
The file contains the distribution of genes for ancient energy conservation complexes (the ATP synthase, Mtr and Rnf) across bacterial classes and archaeal orders.

**Table S7. (separate file) Input data and results of reaction stratification.**
The file contains the food set and reaction set used in the stratification analysis, a list of excluded and added reactions, notes on recoding and the output given by the CatReNet software.

**Table S8. (separate file) Initial set of 424 core metabolic reactions**.
The file contains reaction equations for 424 metabolic reactions required to synthesize amino acids, nucleotides and cofactors from $H_2$, $CO_2$, $NH_3$, $P_i$, $H_2O$ and $H_2S$. They are polarized in the biosynthetic direction, except for irreversible glycolytic reactions. Reactions added in this study to the original dataset from Wimmer *et al.*[18] are marked.

**Table S9. (separate file) List of protein families after sequence-based and structure-based merging and TM-scores for structural alignments.**

The file contains protein families mapped to reaction IDs at each stage of sequence- and structure-based cluster merging. TM-scores obtained for structural alignments are listed.